\documentclass[10pt,letterpaper]{article}
\usepackage[top=0.85in,left=2.75in,footskip=0.75in]{geometry}

\usepackage{amsmath,amssymb}
\usepackage{multirow}

\usepackage{changepage}

\usepackage[utf8x]{inputenc}

\usepackage{textcomp,marvosym}

\usepackage{cite}

\usepackage{nameref,hyperref}

\usepackage[right]{lineno}

\usepackage{microtype}
\DisableLigatures[f]{encoding = *, family = * }

\usepackage[table]{xcolor}

\usepackage{array}

\newcolumntype{+}{!{\vrule width 2pt}}

\newlength\savedwidth

\raggedright
\usepackage[aboveskip=1pt,labelfont=bf,labelsep=period,justification=raggedright,singlelinecheck=off]{caption}

\makeatletter
\renewcommand{\@biblabel}[1]{\quad#1.}
\makeatother

\usepackage{lastpage,fancyhdr,graphicx}
\usepackage{epstopdf}
\fancyheadoffset[L]{2.25in}
\fancyfootoffset[L]{2.25in}
\begin{document}
\vspace*{0.2in}

% Title must be 250 characters or less.
\begin{flushleft}
{\Large
\textbf\newline{Analysis of group evolution prediction in complex networks} % Please use "sentence case" for title and headings (capitalize only the first word in a title (or heading), the first word in a subtitle (or subheading), and any proper nouns).
}
\newline
% Insert author names, affiliations and corresponding author email (do not include titles, positions, or degrees).
\\
Stanisław Saganowski\textsuperscript{1,*},
Piotr Bródka\textsuperscript{1},
Michał Koziarski\textsuperscript{2},
Przemysław Kazienko\textsuperscript{1}
\\
\bigskip
\textbf{1} Department of Computational Intelligence, Faculty of Computer Science and Management, Wrocław University of Science and Technology, Wrocław, Poland
\\
\textbf{2} Department of Electronics, Faculty of Computer Science, Electronics and Telecommunications, AGH University of Science and Technology, Kraków, Poland
\\
\bigskip

% Use the asterisk to denote corresponding authorship and provide email address in note below.
* stanislaw.saganowski@pwr.edu.pl

\end{flushleft}
% Please keep the abstract below 300 words
\section*{Abstract}
In the world, in which acceptance and the identification with social communities are highly desired, the ability to predict the evolution of groups over time appears to be a vital but very complex research problem. Therefore, we propose a new, adaptable, generic, and multistage method for Group Evolution Prediction (GEP) in complex networks, that facilitates reasoning about the future states of the recently discovered groups. The precise GEP modularity enabled us to carry out extensive and versatile empirical studies on many real-world complex / social networks to analyze the impact of numerous setups and parameters like time window type and size, group detection method, evolution chain length, prediction models, etc. 
Additionally, many new predictive features reflecting the group state at a given time have been identified and tested. Some other research problems like enriching learning evolution chains with external data have been analyzed as well.

%\linenumbers

\section*{Introduction}

% influenza
Network science is a very interdisciplinary domain focusing on understanding the relational nature of various real-world phenomena using for that purpose diverse network models. Commonly, networks consist of smaller, more integrated structures called groups, communities, or clusters. In practice, both the groups and whole networks evolve and change their profiles over time. Hence, their analysis demands advanced computational methods to understand and predict their future behavior. For that reason, group evolution prediction is an essential component of computational network science.

One of the domains explored by network science are biological networks\cite{zickenrott2017prediction,barabasi2011network,wu2008network,goh2007human}. Viruses are as old as life on earth. At the same time, they are very young, as they constantly mutate to change their lethal attributes. Influenza, unlike other viruses which are rather stable, evolves much more rapidly\cite{Influenza_rate1:2017,Influenza_rate2:1986} and kills up to one million people worldwide every year\cite{Influenza_kills:2009}. We can try to protect ourselves using vaccines. However, the rate of mutation is too rapid to provide an effective cure. What is more, the development of a new drug requires a huge amount of money and lasts from a few to a dozen or so years.
Despite these difficulties, new drugs are introduced to the market every year. For example, antagonist drugs (also called blockers) are designed to bind to specific receptors to block the disease's ability to attach to these particular receptors, thereby immunizing the body to the disease. Unfortunately, diseases react to drugs and eventually mutate, creating a variety that will bind to other receptors. Therefore, we need methods that will be able to track the evolution of the disease, and based on the history of its mutations, will be able to predict the most likely future mutations. 
To track diseases mutations, we can focus on the group of receptors that it binds to, and observe how such group evolves. Based on the history of changes in the lifetime of this group, we can try to predict what will be the next change. Predicting the direction of the mutation could significantly reduce the amount of time and money needed to study the disease. With such knowledge, we would be able to start preparing the drug in advance and bring it to the market much faster and cheaper.

% diapers
Another area that widely applies network science, especially its branch called social network analysis (SNA), is marketing, in particular advertising\cite{husnain2017impact,antoniadis2016social,guo2016effects,barhemmati2015effects}. Let us imagine that a start-up company invented a new generation of diapers – \textit{Smart Diapers}, which are extra soft, super absorbing, and additionally, can communicate with parents’ smartphones to notify when their change time comes. The company invested very much in their development, therefore, it has a limited budget to advertise the product. The owners decided to introduce the product to discussion groups on the Facebook platform where parents from different countries/cities create and join independent groups to talk about and comment on new products for babies, share general advice about raising children, sell used clothes, etc. Convincing members (parents) of such relevant, targeted groups to use and buy the new diaper product would be much more effective and cheaper than advertising the broader community using expensive TV commercials. Additionally, the word-of-mouth recommendation is commonly believed to be the most powerful marketing tool\cite{Kozinets:2010}.
However, the vital question rises here: which Facebook groups the company should invest in its limited resources, i.e., time and money? In the newly created relatively small groups that might be very active and are expanding fast, or in the larger groups that might be not very active in the nearest future? Which of these groups will be still running or growing in a few weeks/months/years and which one will disappear? That is why the knowledge about the history, current state, and future evolution of groups is crucial at decision making on where to allocate the resources.

In 2007, Palla et al. \cite{palla2007quantifying} have defined the problem of group evolution identification. In the following years, dozens of solutions to this problem have been proposed. One of them was the highly cited GED method \cite{brodka2011tracking}.
Existing surveys describe as many as 12 \cite{saganowski2017community} or even over 60 methods \cite{rossetti2018community}.
All of them are focused on defining possible events in the community life, hence, tracking the historical changes. This, in turn, has led to emerging a new problem – predicting future changes that will occur in the community lifetime. Some of the first methods concerning prediction of some aspects (e.g., determining lifespan) of the group evolution were: (1) Goldberg et al. \cite{goldberg2011tracking} – they focused on predicting the lifespan of evolution for a group; (2) Qin et al. \cite{qin2011evolution} – analyzed dynamic patterns to predict the future behavior of dynamic networks; and (3) Kairam et al. \cite{kairam2012life} – they investigated the possibility of prediction whether a community will grow and survive in the long term.

Note that the methods for tracking group evolution can be also utilized to other similar prediction problems, like link prediction\cite{group_evolution_link_prediction:2017}, churn prediction\cite{group_evolution_churn_prediction:2010}, as well as to understand evolution of software (Unix operating system networks) \cite{evolution_linux:2017} or dynamics of social groups forming at coffee breaks\cite{evolution_coffee_breakes:2014}.

In 2012, we proposed a new concept, in which the historical group changes were utilized to classify the next event in the group's lifetime\cite{Brodka:2012}. In this first trial, we have used only event type and size of the group to describe its state at a given time. Over the next year, we have investigated the concept and adopted it to two methods for tracking group evolution – the GED\cite{brodka2013ged} method and the SGCI method\cite{Gliwa:2012}. This resulted in the first method for group evolution prediction\cite{Gliwa:2013}. It was the predecessor of the GEP (Group Evolution Prediction) method described in this paper. Since then, a few more methods have been proposed. At the end of 2013, İlhan et al. presented their research with several new measures describing the state of the community and a new method for tracking group evolution\cite{Ilhan:2013}. In 2014, Takafolli et al. applied the binary approach to classifying the next change that group will undergo\cite{Takaffoli:2014}. They used 33 measures to describe the state of the community. We have presented new results in 2015, where, apart from new measures, the influence of the length of the history used in the classification was examined\cite{Saganowski:2015}. Later the same year, Diakidis et al. adapted the GED method to conduct their research with 10 measures as predictive features\cite{Diakidis:2015}. In 2016, İlhan et al. presented new results and proposed a method to select measures, which should be the most useful as predictive features for a given data set\cite{Ilhan:2016}. More recently, Pavlopoulou et al. used 19 measures already validated in other works and studied whether employing the temporal features on top of the structural ones improves prediction, as well as what is the impact of using a different number of historical community states on the prediction quality\cite{Pavlopoulou:2017predicting}.

Unfortunately, all of the methods proposed to this day have some drawbacks (see the Comparison with other methods section) and have been designed to solve a particular problem, hence, their application area is rather narrow. 
Therefore, in this paper, a new generic and comprehensive method to predict the future behavior of the groups, based on their historical structural changes as well as experienced events, is proposed, evaluated and discussed.

Some of the contributions of this work are: decomposing the group evolution prediction problem, proposing and extensively evaluating the modular method that can be applied to any dynamic network data, proposing new predictive features, performing the features' ranking, proposing a new concept of data set enriching, initial evaluation of the transfer learning technique, an example and discussion on the concept drift problem in group evolution prediction, reviewing all proposed methods in the field.

\section*{Methods}

\subsection*{Decomposition of the group evolution prediction problem }

The crucial matter in developing the modular method predicting group evolution, called \textit{GEP}, was the identification and separation of the components of the entire group evolution prediction problem. The appropriate problem decomposition and information flow between particular components (dependencies) are depicted in Eq~\ref{eq:decomposition} and Fig.~\ref{fig:GEP_method}.

\begin{equation} \label{eq:decomposition}
IS\xrightarrow[S_1]{TWT}TW\xrightarrow[S_2]{NT}TSN\xrightarrow[S_3]{CDM}G\xrightarrow[S_4]{CETM}EC\xrightarrow[S_5]{FE}PF\xrightarrow[S_6]{classification(CH)}Q
\end{equation}

The data from the input stream $IS$ is divided into time windows $TW$ using the time window type definition $TWT$. For each time window $TW$, a complex/social network is created using the network type definition $NT$, resulting in the temporal complex/social network $TSN$. Within each time window $TW$ in $TSN$, some groups $G$ are identified using a community detection method $CDM$. Next, similar and consecutive groups are matched using a community evolution tracking method $CETM$, as well as the transition is labeled with an event type out of the set of possible changes $CH$. The matched groups are combined into evolution chains $EC$ that may consist of many successive changes. For each community state in $EC$, the feature extraction process $FE$ is applied in order to obtain a set of predictive features $PF$ describing the community state at a given time. Using features $PF$ in the form of a vector representing each evolution chain $EC$, classification of possible changes $CH$ is performed. The classification task (stage $S_6$) is to learn and finally label the next change(s) in community lifetime. The output of the classification process is a set of classification quality (performance) measures $Q$, for example, F-measure, accuracy, precision, or recall.
The identified components were converted into six stages $S_1$-$S_6$ of the GEP method, Fig.~\ref{fig:GEP_method}.

\subsection*{GEP method}

\begin{figure}[!ht]
\centering
\includegraphics[width=\linewidth]{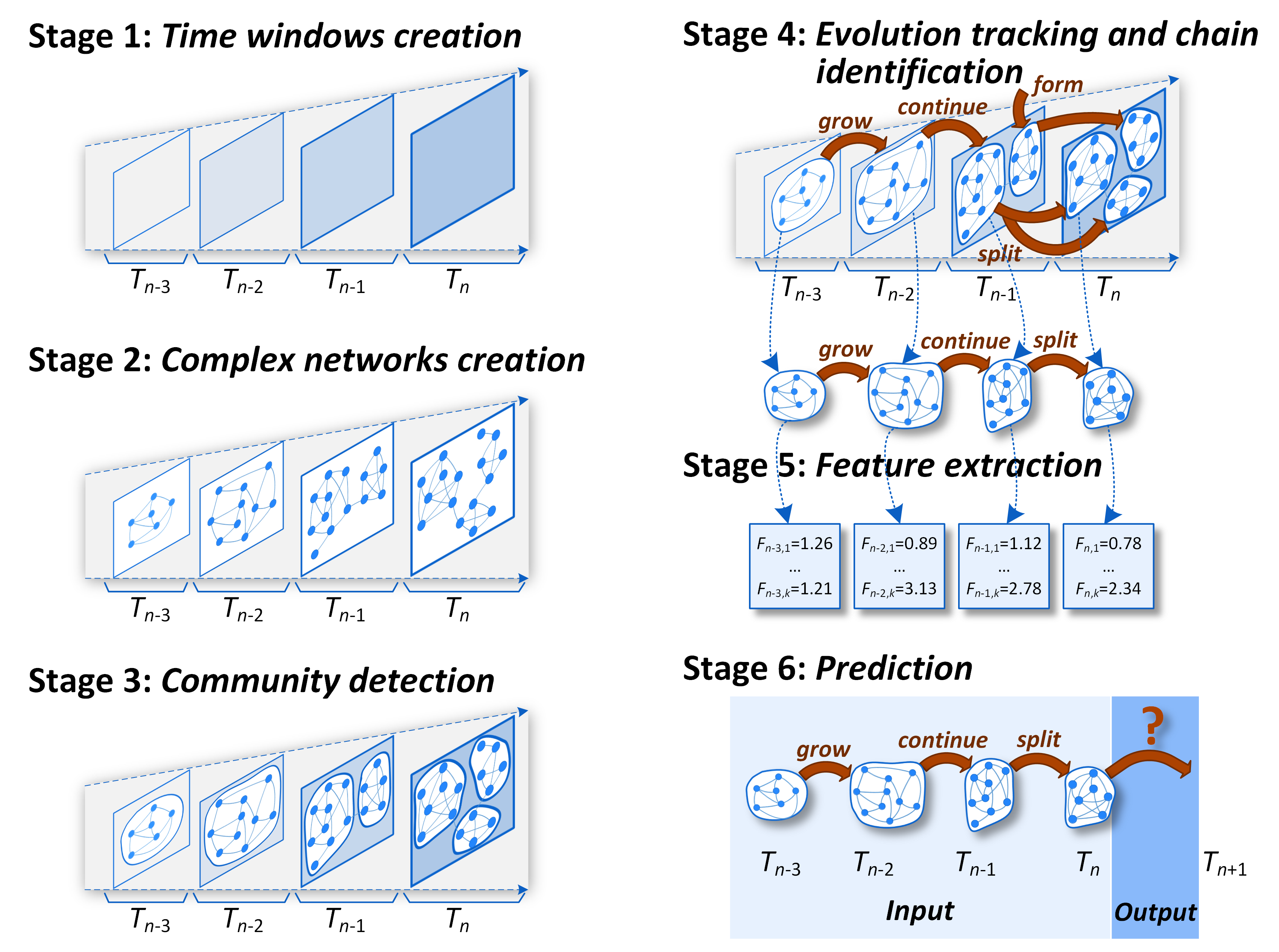}
\caption{\textbf{The concept of the GEP method.} \textbf{Stage 1:} Data set is divided into time windows. \textbf{Stage 2:} A complex network for each time window is created. \textbf{Stage 3:} Groups are extracted within each time window using any community detection method. \textbf{Stage 4:} The evolution of communities is tracked with any group evolution tracking method, and the evolution chains are created. \textbf{Stage 5:} Features describing the previous group profile such as size, density, cohesion, etc. are calculated to capture community state at a given time. \textbf{Stage 6:} Supervised machine learning approach is applied to learn and predict the forthcoming event in the group's lifetime.}
\label{fig:GEP_method}
\end{figure}

The GEP framework consists of six main stages (Fig.~\ref{fig:GEP_method}): (1) time window definition, (2) complex network extraction for the defined periods, (3) community detection in periods, (4) group evolution tracking, (5) evolution chain identification for communities together with feature extraction and computation for each chain and (6) classification, containing classification model learning and testing. Each of them can be implemented by means of different methods and approaches depending on research need and prerequisites, e.g., complexity level. The formal definition of the GEP method is as follows:

\newtheorem{gep}{Definition}
\begin{gep}
The GEP method is defined as an octuple $<IS, S_1, S_2, S_3, S_4, S_5, S_6, Q>$, where:
\newline$IS$ is an input stream of activities, e.g., phone calls, linking two actors (network nodes) $x, y$ at time $t_i$;
\newline$S_1$ is a set of considered time windows of the given type $TWT$;
\newline$S_2$ is a set of considered approaches to temporal complex / social network $TSN$ creation from $IS$ using time window definitions from $S_1$;
\newline$S_3$ is a set of considered approaches to community detection methods $CDM$ for each time window in $TSN$ from $S_2$;
\newline$S_4$ is a set of considered approaches to tracking community evolution methods $CETM$ for communities from $S_3$;
\newline$S_5$ is a set of considered approaches to feature extraction for evolution chains from $S_4$;
\newline$S_6$ is a set of considered approaches to classification, including learning, training, validating, undersampling, oversampling, and feature selection techniques;
\newline$Q$ is a set of considered classification quality measures, for example, F-measure, accuracy, precision, recall, estimated based on the classification results from $S_6$.
\end{gep}
The methods enumerated especially in $S_1$, $S_3$, $S_4$, $S_6$ also include the space / set of their parameters. 

The output of one stage $S_i$ is the input for the next stage $S_{i+1}$, e.g., communities detected in $S_3$ are used to discover their evolution in $S_4$. All these stages, together with parameters of the methods used, are more in-depth described in \nameref{SI_File}. They also require an appropriate definition of data structures to facilitate hassle-free implementation.

\subsection*{CPM method}
The Clique Percolation Method (CPM) proposed by Palla et al.\cite{Palla:2005CPM}  is the most widely used algorithm for extracting overlapping communities. The CPM method works locally, and its primary idea assumes that the internal edges of a group have a tendency to form cliques as a result of high density between them. Oppositely, the edges connecting different communities are unlikely to form cliques. A complete graph with $k$ members is called k-clique. Two k-cliques are treated as adjoining if a number of shared members is $k$–1. Lastly, a k-clique community is the graph achieved by the union of all adjoining k-cliques. Such an assumption is made to represent the fact that it is a crucial feature of a group that its nodes can be attained through densely joint subsets of nodes.

\subsection*{Infomap method}
The Infomap method proposed by Rosvall and Bergstrom\cite{Rosvall:2008infomap} uses the information-theoretic approach to cluster nodes within a network. It focuses on information diffusion across the graph and compression of the information flow description obtained from a random walker, which is chosen as a mean of information diffusion. Infomap changes the problem of finding the best cluster structure into finding the partition with the minimum description length of an infinite random walk. It follows the intuitive idea that if the community structure is present, the random walker will spend more time inside the community because of its higher edges density. It means that the transition to another cluster will be less likely.

\subsection*{GED method}
The Group Evolution Discovery (GED) method\cite{brodka2013ged} is one of the best methods for tracking community evolution\cite{he2017comparative}. It uses inclusion measure to match similar communities from neighboring time windows. This measure takes into account both the quantity and quality of the group members. The quantity is reflected by the first part of the inclusion measure, i.e., what portion of the members from group $G_1$ also belongs to group $G_2$. The quality is expressed by the second part of the inclusion measure, namely, what contribution of important members from group $G_1$ is in $G_2$. It provides a balance between the groups that contain many of the less important members and groups with only few but key members. The inclusion measure and the group size determine the type of community change. The authors defined seven possible event types: forming, dissolving, continuing, growing, shrinking, merging, and splitting. The method can work with any community detection method and with any group similarity measure, thus, providing great flexibility.

\subsection*{İlhan et al. method}
The İlhan et al. method\cite{Ilhan:2016} works with the disjoint type of communities and utilizes the function by Hopcroft et al.\cite{Hopcroft:2004} to calculate the similarity between two communities. The event types that can occur in the community lifetime and also the classes being classified are: survive, growth, shrink, merge, split, and dissolve. The measures used as predictive features are divided into two categories: structural and temporal community measures. In total, nine features per timeframe are used, i.e., number of nodes and edges, intra and inter measure of community edges, betweenness, degree, conductance, aging, and activeness. If one calculates four network measures beforehand (average path length, betweenness, clustering coefficient, embeddedness), the method can also identify features that should be the most prominent for a given network profile.

% Results and Discussion can be combined.
\section*{Results}

Suitable decomposing the problem of group evolution prediction (see the Methods section and Fig.~\ref{fig:GEP_method}) was crucial in solving the problem. It allowed to analyze distinct phases of the process and to propose multiple solutions for each phase. The GEP method was extensively analyzed on fifteen real-world data sets (see \nameref{SI_File} for their profiles), for which more than 1,000 different temporal networks were created, and in total, more than 5,000,000 individual classification tasks were performed. However, to keep the article clear and concise, only selected results are presented for each stage.

\subsection*{Stage 1: Time windows creation}
At first, the data is divided into time windows. Three main approaches can be considered in this context: (1) equal length periods – the events and relations are segmented based on their timestamp; (2) the same number of relations in each time window; (3) the arbitrary division, based on the data context. Additionally, the type and size of time windows have to be decided, which may be a challenging task. There are three most common types of time windows: disjoint, overlapping, and increasing.

A proper choice of the time window type and size has a direct impact on the following GEP stages, especially on the number of evolution chains discovered by the tracking method (Stage 4). If relations between individuals in a data set have a tendency to change rapidly, then disjoint time windows would be a poor choice since there may not be too many relations lasting between two consecutive time windows. As a result, the tracking method will not provide any events (Stage 4), so there will be no input to a classifier resulting in no event to predict (Stage 6). The too large size of the time window, in turn, might lose some information about community changes that occurred in the meantime.

So far, there is no formula which determines the right type and size of the time window, but a few guidelines can be provided based on our extensive experiments:
\begin{itemize}
\item If the network is sparse or changes rapidly, the overlapping time window should be used. Usually, the offset equal to 30\% of the time window size is enough to obtain a reasonable number of events between the consecutive time windows;
\item The time window type and size should be adjusted to the context of the given data set, e.g., the co-authorship network, referring to researchers who often publish only once a year, should evolve smoothly with the 1-year disjoint time windows;
\item If the persistent groups are the goal of analyses, the increasing time window should be utilized, as it provides mostly the continuing and growing events;
\item If relations between individual nodes are recurrent and the network is rather dense, one may try using disjoint time windows to lower the computational cost;
\item It is acceptable and even preferable to repeat the selection of the time window type and size several times to see which approach yields the best results.
\end{itemize}
The most common choice in our studies was the overlapping time windows with the offset between 30\% - 50\% of their size.

\subsection*{Stage 2: Formation of networks}
The parameters that can be adjusted at the creation of networks for each time window is the set of edge attributes, in particular, their weights and direction. The weighted/unweighted, as well as directed/undirected profile of the network, did not yield a significant impact on computational complexity nor classification accuracy. Some community detection methods, however, may be incompatible with the networks of particular characteristics or may ignore some attributes, e.g., weights. The CPM\cite{Palla:2005CPM} and Infomap\cite{Rosvall:2008infomap} methods, used in the experimental studies, are capable of handling the most important network attributes. 

\subsection*{Stage 3: Community detection}
\begin{figure}[!ht]
\begin{adjustwidth}{-2.25in}{0in}
\centering
\includegraphics[width=\linewidth]{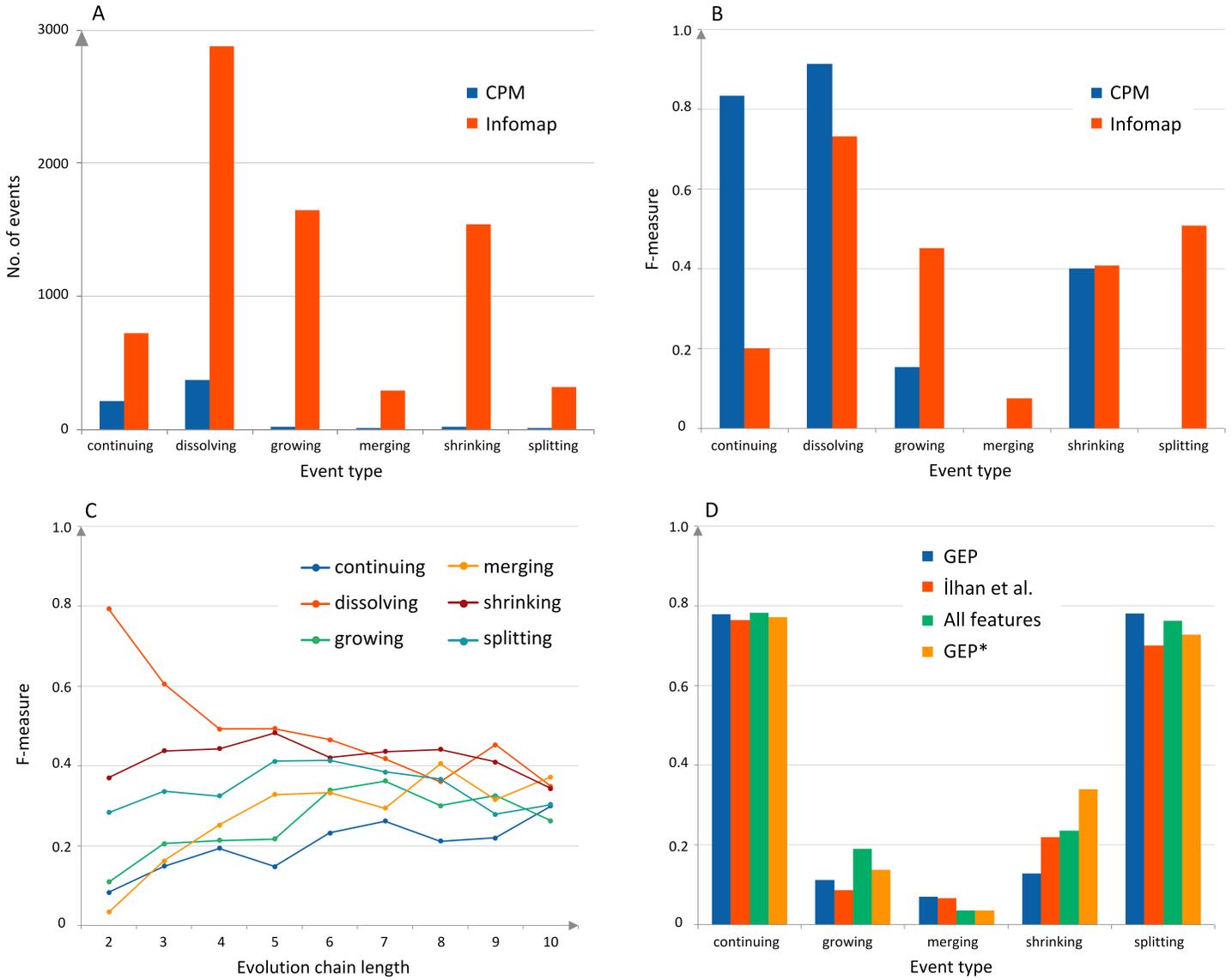}
\caption{\textbf{(A) CPM vs. Infomap.} The number of events tracked with the GED method for groups obtained with two different community detection methods applied to the Digg data set. The CPM method leaves out even 97\% of nodes that do not belong to any clique, hence the small number of groups and events. \textbf{(B) CPM vs. Infomap.} 
The F-measure values achieved for the events presented in Fig.~\ref{fig:Chain CPMvsInfomap GEPvsIlhan}A. The results reflect the distribution of events. \textbf{(C) Chain length.} The F-measure values for different lengths of the evolution chains for the Facebook data set. For most of the events, the F-measure value was increasing with the increase of the chain length up to 6 or even 7 states (the continuing and growing events). Beyond that point, the number of evolution chains of the particular types dropped below 50 which was insufficient to train the classifier properly;  \textbf{(D) GEP vs. Ilhan et al.} The F-measure values for the 9-state evolution chains obtained from the Slashdot data set with the different set of predictive features: only from the GEP method (GEP) - see \nameref{SI_File}, from the İlhan et al. method, combined from both GEP and İlhan et al. methods (All features), and from the GEP method, but only for the last 3 states out of all 9 states (GEP*). The GEP* and "All features" scenarios achieved slightly better overall scores.}
\label{fig:Chain CPMvsInfomap GEPvsIlhan}
\end{adjustwidth}
\end{figure}

Some community detection methods can produce both disjoint and overlapping communities, but there are only a few methods for tracking the evolution (Stage 4) that can deal with the overlapping groups. Overall, the methods extracting disjoint communities perform faster than the ones providing overlapping groups. In some extreme cases, when the network is very large, the CPM method is unable to extract groups due to its enormous memory requirements. It is hard to compare two types of the grouping methods in terms of their impact on the classification accuracy, as each type of clustering delivers a different set of communities resulting in a different distribution of evolution events.
Besides, the profile of the groups may be diverse, e.g., networks grouped with the CPM method tend to have a single giant component with many small overlapping groups alongside. This method also inclines to leave out nodes that do not belong to any clique, thus, excluding them from further consideration. If the network is sparse, a major fraction of the network may be omitted. In the most extreme case, the CPM method neglected even as many as 97\% of network nodes, what resulted in a deficient number of communities and evolutions (Fig.~\ref{fig:Chain CPMvsInfomap GEPvsIlhan}A), and eventually in very low classification accuracy, Fig.~\ref{fig:Chain CPMvsInfomap GEPvsIlhan}B. At the same time, the Infomap method performed very well, identifying a large number of communities.
Furthermore, the overlapping groups are likely to generate more merging and splitting events in Stage 4, since there are plenty of similar and overlapping communities in the consecutive time windows. On the other hand, the Infomap method tends to produce many communities having only 2 or 3 nodes.
In general, while considering which type of grouping method to use the data context should be a crucial factor.

\subsection*{Stage 4: Stepwise evolution tracking and chain identification}
Regardless of the method, tracking the evolution of community is a computationally demanding task. The method has to iterate over all time windows and compare all the communities in order to detect similar ones. Although the methods for tracking group evolution can be very distinct, especially while defining the possible event types, our earlier study showed that the selection of the method has no significant impact on classification accuracy\cite{Saganowski:2015}. In this evaluation, we use the GED method \cite{brodka2013ged} since, in the last evaluation of existing community evolution tracking method, it was selected as the one giving the most satisfying results\cite{he2017comparative}. 

The parameters of the selected method might influence the classification results, e.g., the alpha and beta parameters of the GED method have a direct impact on the number of evolution events discovered – the lower the threshold, the more events obtained (see \nameref{SI_File} for details). In the experimental studies, the most common value for the alpha and beta parameters was 50\%. If the network is dense and relations are recurrent, the alpha and beta might be even increased to 70\%. On the other hand, when the method provides a small number of the evolution events, the alpha and beta should be reduced to, e.g., 30\%. 
Apart from the selection of the evolution tracking method, the length of the evolution chain has to be decided. The longer the evolution chain, the more predictive features for the classifier in Stage 6, hence, the higher computational complexity. Nevertheless, the results presented in Fig.~\ref{fig:Chain CPMvsInfomap GEPvsIlhan}C revealed that it is worth dedicating some more time and resources to extract longer chains since it can boost classification accuracy. The overall score achieved with the evolution chains containing six community states was 32\% higher than the results achieved with shorter 2-state chains. In case of limited time or resources, the chains with the length of 2-3 states should be reasonably good.

\subsection*{Stage 5: Feature extraction}
\begin{figure}[!ht]
\begin{adjustwidth}{-2.25in}{0in}
\centering
\includegraphics[width=\linewidth]{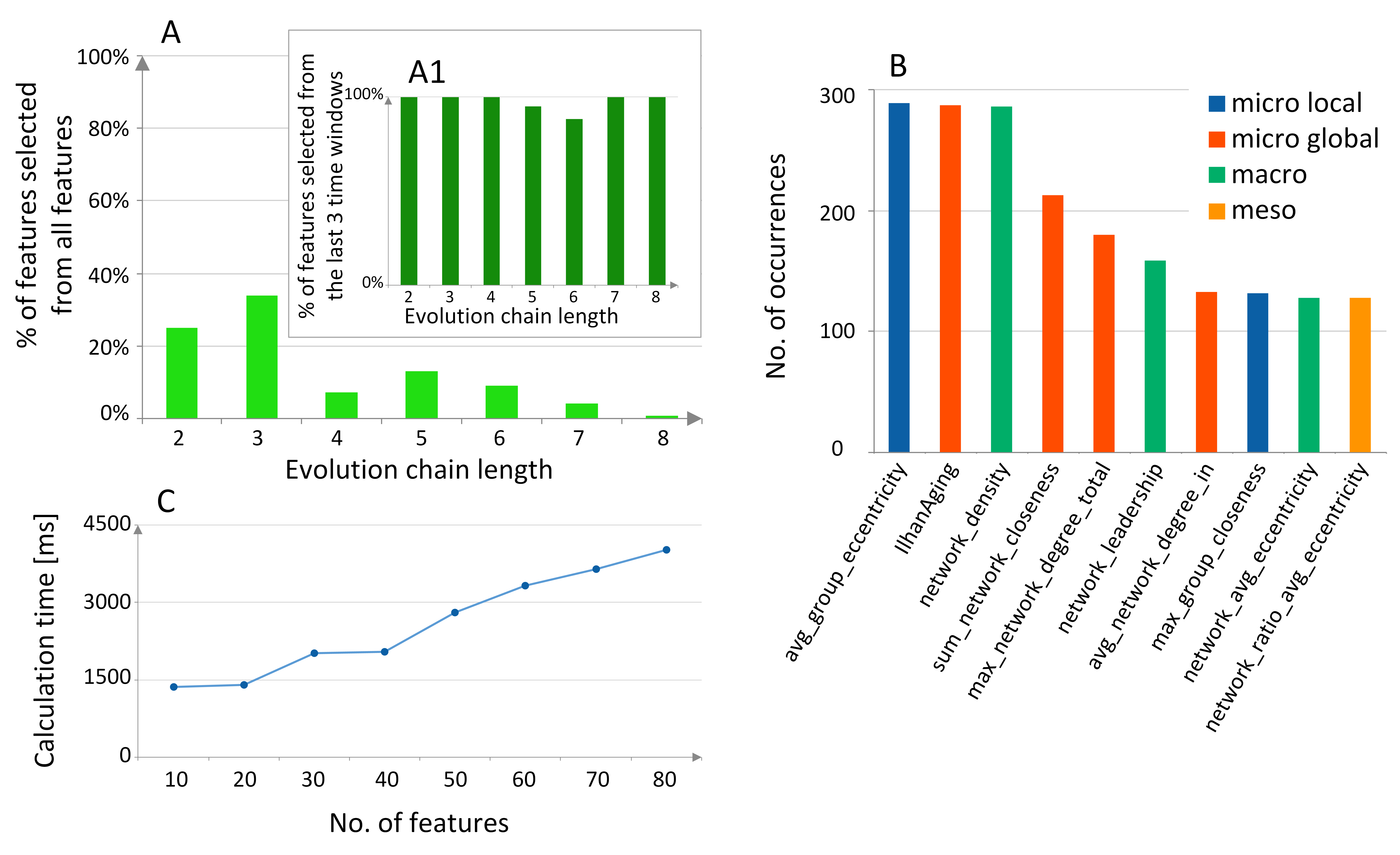}
\caption{\textbf{(A) Feature selection.} Important features selection obtained by the Backward Feature Elimination for the DBLP data set. The total number of features increases with every state by 91, e.g., the 3-state evolution chain has 91*3=273 features in total, out of which 34\% were selected as prominent. \textbf{(A1)} Features selected only from those related to the last 3 time windows. \textbf{(B) Feature ranking.} The most frequently selected features for the 1-state evolution chains. All kinds of information are important to achieve a satisfactory prediction; microscopic features are focused on nodes, mesoscopic on groups, and macroscopic on entire network parameters. The ranking obtained by analyzing eight data sets and repeating feature selection 1000 times. \textbf{(C) Computational efficiency.} The time required to train a single Random Forest classifier in relation to the number of descriptive features used as the input data. The results obtained for the IrvineMessages data set.}
\label{fig:Features selection}
\end{adjustwidth}
\end{figure}

In order to predict the future evolution of the group, we need to describe its recent and historical states by means of predictive features. Based on these features and previous evolutionary changes used to learn the model, we are able to forecast the next changes. The crucial features that are at our disposal are structural network measures computed for the previous group states. 
Calculation of all measures may be a very demanding task since they need to be evaluated for every community state in the evolution chain. Additionally, some measures, e.g., betweenness centrality, require finding all shortest paths for each pair of nodes in the community or network. The experiments revealed, Fig.~\ref{fig:Chain CPMvsInfomap GEPvsIlhan}D, Fig.~\ref{fig:Features selection} that the set of predictive features has a significant impact on classification accuracy, as they are used to build the classification model, see also \nameref{SI_File}, Feature Selection section. Therefore, it is highly recommended to compute as many predictive features as possible to deliver to the classifier a wide variety of descriptions to choose from. 

To significantly enhance the already existing approaches, many new predictive features are proposed in this paper (see \nameref{SI_File}, Predictive Features section). We have clustered structural features into three general types: (1)~\textit{microscopic} -- calculated for individual nodes, e.g., node degree, (2) \textit{mesoscopic} -- quantifying single groups, e.g., group size - no. of nodes, and (3) \textit{macroscopic} -- describing the whole network, e.g., network density.  Mesoscopic features also include normalized group measures like the group size divided by the network size.
Besides, node-based (microscopic) measures can be aggregated (usually averaged) at either \textit{local} (group) or \textit{global} (network) level resulting in \textit{microscopic local} or \textit{microscopic global} features, respectively. 

All computed features were thoroughly evaluated in terms of usefulness for the classifier and rankings of the most prominent features were built, see \nameref{SI_File}, Feature Selection section, especially Tab.~5-9. For the evolution chains of a variable length, different rankings were obtained. For the shortest 1-state evolution chains, only macroscopic (network) features were helpful, which may result from the fact that communities with a short history are considered unstable and vulnerable to the environment they are a part of. For the evolution chains with the increasing time windows, the features describing the local structure, especially the centrality- and distance-based measures, were more informative for the classifier, as the changes between the consecutive increasing time windows were delicate and occurred at the microscopic rather than macroscopic level. The neighborhood-based features were among the most valuable features for the longest 8- and 9-state chains, which lead to believe that for the long-lasting communities, the relations with their surroundings are a better predictor of the forthcoming change than, e.g., the macroscopic features. In general, the variations of the eigenvector-, eccentricity-, and closeness-based features were present in most of the selective rankings, which suggests that centrality- and distance-based measures obtained on the node level are the most prominent ones. Hence, in case of limited computational capacity, these features should be respected before any other. However, out of all features considered by the classifiers, the Backward Feature Elimination selected only up to 34\% of them as prominent, i.e., used by the classifier to make a decision, Fig.~\ref{fig:Features selection}A.

Additionally, it turned out that usually over 90\% of the selected prominent features were obtained from the last three community states, Fig.~\ref{fig:Features selection}A1. For example, when the evolution chain length was 8, and the next change was classified, all the prominent features were from the 8th, 7th, and 6th group profiles. It means that the most recent history of the community has the most significant impact on its next change. This is an extremely useful conclusion if one has limited computational capabilities and cannot calculate community profiles for all states or does not possess data about older history. The number of features has a direct impact on the duration of the entire learning process, Fig.~\ref{fig:Features selection}C.

\subsection*{Stage 6: Prediction}

In the last stage, the machine learning techniques, such as oversampling, undersampling, feature selection, and first of all, model training and adjustment are applied to achieve the highest possible prediction quality. The common problem with the training data is an imbalanced distribution of output classes, Fig.~\ref{fig:Chain CPMvsInfomap GEPvsIlhan}A. In extreme cases, when one class greatly dominates over the other ones, a trained model tends to assign the dominant class to most observations. Then, the solution is to apply additional preprocessing techniques like oversampling and undersampling to generate additional observations or to filter out predominant ones, thus providing a distribution closer to flat. 
Another common problem is overfitting the classifier by providing too many features or observations. In order to prevent from such case, feature elimination technique may be applied, which unfortunately is very expensive in terms of computational complexity. 

\begin{figure}[!ht]
\begin{adjustwidth}{-2.25in}{0in}
\centering
\includegraphics[width=\linewidth]{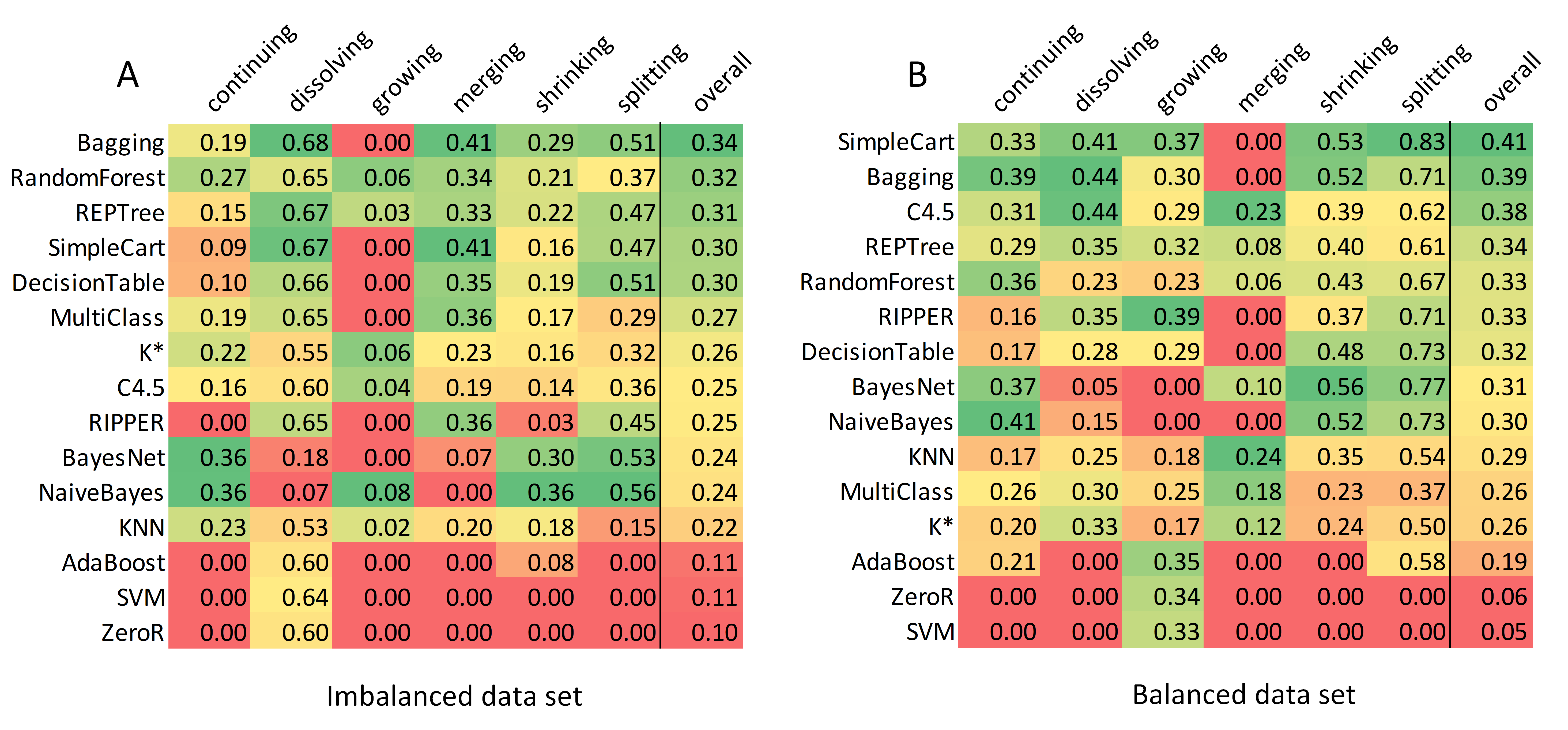}
\caption{\textbf{The rankings of classifiers.} The heat-maps of the F-measure results for the 1-state evolution chains obtained from the Twitter data set. Classifiers are ordered by the overall score. The Bagging classifier and the SimpleCart classifier achieved the highest overall scores but failed to predict the growing and the merging events. Therefore, the tree-based classifiers are the best choice as all the events are successfully classified and the overall score is insignificantly lower.}
\label{fig:Ranking of classifiers}
\end{adjustwidth}
\end{figure}

Additionally, the proper classifier should be selected, and its parameters need to be accordingly adjusted. In the experimental study, fifteen different classifiers were compared in terms of the classification accuracy, Fig.~\ref{fig:Ranking of classifiers}. The tree-based classifiers and meta-classifiers (equipped with decision trees) performed best. Many classifiers could not efficiently handle imbalanced data, so the undersampling and oversampling techniques were applied, resulting in notably better prediction quality, Fig.~\ref{fig:Ranking of classifiers}B. On the balanced data set, a classifier focuses on the predictive features computed for the community states instead of focusing on the event distribution.

The Friedman statistical test~\cite{Friedman:1937} with the Shaffer post-hoc multiple comparisons~\cite{Shaffer:1986} was performed to obtain rankings of classifiers on the imbalanced and balanced data sets (cf. \nameref{SI_File}, Tab.~10). In both cases, the Bagging classifier (with the REPTree classifier) was the winner, and the Random Forest classifier was ranked second. What is essential, the p-values confirmed that the results were statistically significant.

Furthermore, classifiers often have their parameters to tune them accordingly, which  can substantially affect the classification accuracy, cf. \nameref{SI_File} for detailed discussion. For example, the logarithmic correlations were observed between the number of bagging iterations for the Bagging classifier and the average F-measure value, as well as between F-measure and the number of generated trees by the Random Forest classifier. The results prove that the process of adjusting the classifier parameters should always be performed, as long as the computational time and resources are available.

\subsection*{Comparison with other methods}
%porownanie z innymi metodami

The GEP method was compared to other approaches. The existing methods for group evolution prediction were additionally analyzed, and many of their drawbacks have been identified. The most severe were: a narrow application area, methodological issues (e.g., inappropriate computation of the conditional probability), insufficient validation of the methods (e.g., a single sampling into two folds instead of the 10-fold cross-validation), superficial descriptions of the methods and conducted experiments (often insufficient to repeat and validate the experiments), and lack or unreliable comparisons with other methods.

Despite GEP is so flexible and has so many options, it is competitive with other approaches, designed to deal with a specific problem or data set. 
For example, a special version of the GEP method, in which only features from the last three states (out of all 8 or 9 states) were used as an input for the classifier, performed noticeably better than the method by İlhan et al.\cite{Ilhan:2016}, Fig.~\ref{fig:Chain CPMvsInfomap GEPvsIlhan}D.

After all, it needs to be emphasized that none of the existing methods is as adjustable and versatile as the GEP method.

\section*{Discussion}

Across its six stages, the GEP method utilizes various approaches, methods, and techniques, which can be adjusted with respect to a given data set and a particular study purpose. These approaches, methods, and techniques are considered as the GEP method parameters. To provide a concise summary of their impact on overall computational complexity, and first of all on the final classification accuracy, the crucial parameters were listed in Tab.~\ref{tab:parametersInfluence} and discussed throughout the article.

\begin{table}[!ht]
\begin{adjustwidth}{-2.25in}{0in}
\centering
\caption{{\bf The GEP method parameters and their impact on computational complexity and classification accuracy.}}
\begin{tabular}{p{2.3cm}|p{4cm}p{5.2cm}p{2.3cm}p{2cm}}
\hline
Parameter group & Parameter & Parameter value & Impact on computational complexity & Impact on classification accuracy \\ \hline
\multirow{3}{*}{time window} & window division & timestamp / relations count / arbitrary & none & low \\ \cline{2-5}
 & window size & time unit or number of relations & medium & low \\ \cline{2-5}
 & window type & disjoint / overlapping / increasing & medium & medium \\ \hline 
network type & edge attributes & directed / undirected, weighted / unweighted & low & low \\ \hline
\multirow{5}{*}{methods} & group type & disjoint / overlapping & medium & low \\  \cline{2-5}
 & grouping method & a method & high & medium \\  \cline{2-5}
 & tracking method & a method & medium & low \\  \cline{2-5}
 & GED alpha and beta & (10\%, 100\%] & none & low \\  \cline{2-5}
 & GED social position measure & a measure & medium & low \\ \hline
 \multirow{2}{*}{classification} & classifier used & a classifier & medium & medium \\  \cline{2-5}
 & machine learning techniques & undersampling, oversampling, feature selection & high & high \\ \hline
 \multirow{2}{*}{other} & evolution chain length & number of community states & medium & medium \\  \cline{2-5}
 & predictive features & a set of features & high & high \\ \hline
\end{tabular}
\label{tab:parametersInfluence}
\end{adjustwidth}
\end{table}

Many different classifiers were evaluated on various data sets. The tree-based classifiers and meta-classifiers (equipped with decision trees) performed best. Many classifiers could not handle imbalanced data sets, so the undersampling and oversampling techniques were applied. Balancing data sets notably improved the results confirming the usefulness of the undersampling and oversampling methods. The experimental studies showed that adjusting the classifier parameters can significantly improve classification accuracy. The logarithmic correlations were observed between the number of bagging iterations in Bagging classifier and the average F-measure value, as well as between the number of generated trees by the RandomForest classifier and the average F-measure value. The confidence factor parameter of the J48 classifier was found also correlated with the average F-measure value. The maximum improvement in average F-measure value achieved by adjusting the classifier parameter was 17\%, and it was obtained by increasing the number of generated trees by the RandomForest classifier. The results prove that the process of adjusting the classifier parameters should always be performed, as long as the computational time and resources are not limited.

%transfer learning
The GEP method enables us to consider different new scenarios, which are hardly available without this generative framework like transfer learning, class balancing by adding external data, or decreasing the concept drift effect. 

The transfer learning technique was adapted to the problem of group evolution prediction for the first time in this field. Its main idea is to learn the classification model on one data set and test it on another one. Such an attempt was quite successful, and the preliminary results were satisfactory. The key to success is finding a data set with a likewise profile. Moreover, in some cases, learning the transferred model on the balanced data set can boost the classification quality for the data set to which the model is adapted. The initial experiments also suggest that the underlying similarity of two data sets (e.g., the same habits of actors or ideally the same set of actors) can help to create a model that if transferred can outperform the primary model built for a given data set.

\begin{figure}[!ht]
\begin{adjustwidth}{-2.25in}{0in}
\centering
\includegraphics[width=\linewidth]{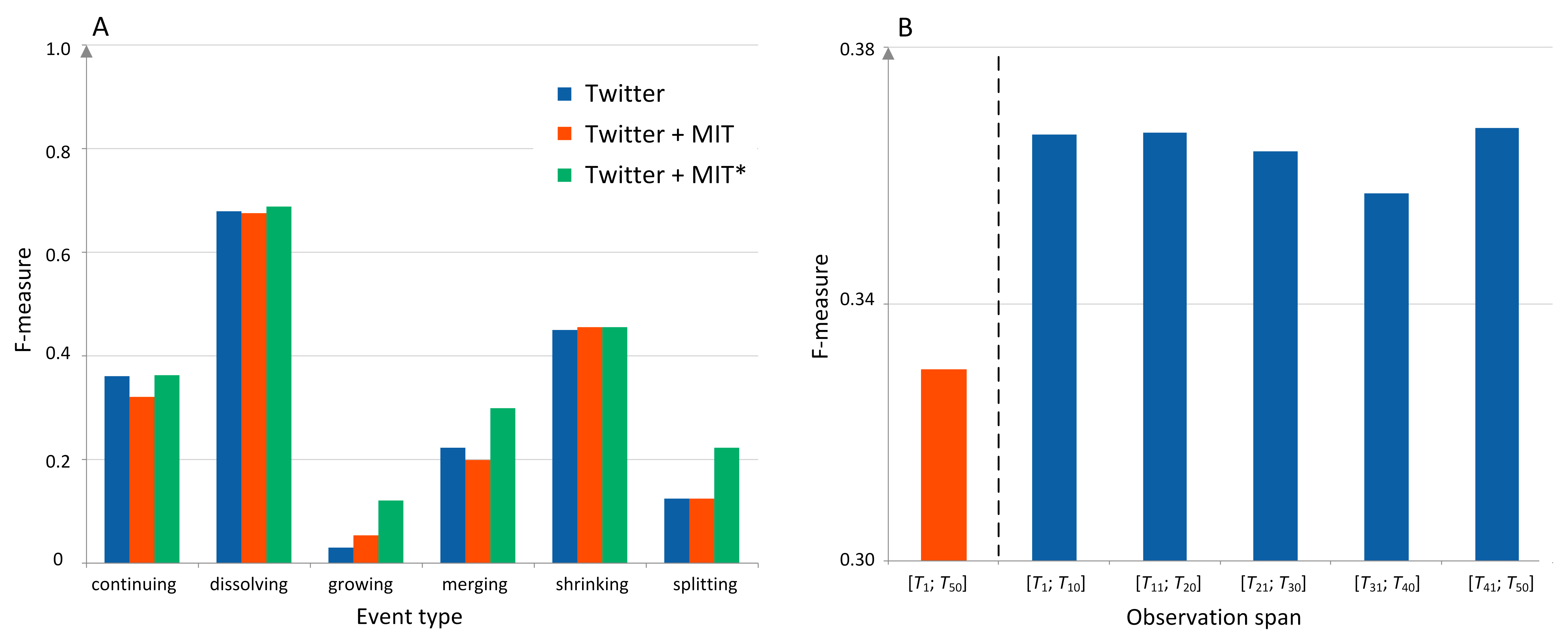}
\caption{\textbf{Application of the GEP method.} \textbf{(A) Enriching the classification model} by partially balancing the original training set (Twitter) with extra evolution chains taken from another full external data set (MIT) or with chains from only selected event types, i.e. growing, merging and splitting (MIT*); chains with these classes were the worst classified events for the original Twitter data -- they had the lowest F-measure values. The results for these selectively enriched event types were significantly improved without worsening classification for other classes (green vs. blue bars).  Data enriching was performed only for learning, not for testing. \textbf{(B) Concept drift.} Classification quality for the Facebook data from one longer period $T_1-T_{50}$ (the red bar); alternatively, the data was split into five smaller periods and separate classification models were built to catch concept drift phenomena between periods (blue bars). Independent models learned for smaller periods are better adapted to the changing environments.}
\label{fig:applications}
\end{adjustwidth}
\end{figure}

%balansowanie zbioru treningowego obserwacjami z innych zbiorow
Very promising results, although at an early stage, were achieved at enriching the learning phase of the classification model with additional evolution chains from a different data set. By partially balancing the original training set with extra evolution chains from another external data set, it was possible to improve the model and thus produce better results for minority classes, without affecting the outcome for the dominating classes, Fig.~\ref{fig:applications}A. This phenomenon is especially important because the existing techniques of balancing a data set always affect the classification results of the dominating classes.

%krotszy observation span
Another way to enhance the classification model, initially considered, is an appropriate selection of the observation time span to reduce the effect of non-stationarity of data -- a.k.a concept drift. Our preliminary research shows that for a network spanning over a long period or changing rapidly, updating the classification model every once in a while might improve the results, as the model reflects the current characteristics of the network in the better and more up to date way, Fig.~\ref{fig:applications}B. Nevertheless, in order to rebuild the model every now and then, the number of observations (evolution chains) extracted from such shorter time span must be high enough.

The GEP framework can be applied to any dynamic network data, i.e., to any complex network changing over time. In this paper, we have explored popular social network data, see Table 2 in the Supporting information section. However, the entire GEP method, its stages and component solutions may be used for diverse complex networks \cite{Barzel2013, BOCCALETTI2006} like evolving clusters of web pages \cite{Dezso2006}, co-citation and bibliographic coupling networks extracted from citations between scientific papers \cite{kessler1963, small1973}, biological and medical networks \cite{Bode2007, Barabasi2011}, linguistic networks linking word meanings - WordNets \cite{Bartusiak2019}, multimedia networks \cite{Indyk2013} and many more.

\section*{Conclusion}

The main subject studied in this paper is group evolution prediction in social/complex networks. Its primary goal is to foresee a change like shrinking, growing, splitting, merging, or dissolving that the recently existing community will experience in the nearest future. To be able to perform any prediction, the most common approach is to process a temporal complex network $TSN$ extracted from the stream of user activity traces. Communities and their changes are identified and predicted within such $TSN$. However, the existing methods are often limited to operate on a particular data set or to solve a specific problem, which makes them useful only in a particular and narrow domain.

Therefore, a new generic method called Group Evolution Prediction (GEP) has been proposed in this paper. The GEP method has a modular structure, which makes it very flexible and allows us to successfully apply it to any data set and under any specific requirements. The method consists of several stages; each of them involves a suitable selection of methods, algorithms, and attributes -- the GEP method parameters. 
The evaluation process of the GEP method included: (1) analysis of numerous parameters (time window type and size, community detection method, evolution chain length, classifier used, set of features, and more), (2) comparative analysis against other existing methods, (3) adaptation of the transfer learning concept to group evolution prediction, (4) enriching the classification model with evolution chains from a different data set, and (5) enhancing the classification model with a more appropriate training set.

Regarding the time window types and sizes, the main finding is that for rapidly changing or sparse social networks a shorter overlapping time windows (in relation to the context of the data) are a better choice than longer or disjoint periods. On the contrary, if relations between individuals are recurrent and the network is rather dense, one may try disjoint time windows to obtain more concise results and to lower the computational cost. If long-lasting, persistent communities are the goal, then the increasing type of time window is the best choice as it generates a high number of the continuing, growing, and shrinking events.

Two most commonly used community detection approaches were analyzed: the CPM method detecting the overlapping communities, and the Infomap method identifying the disjoint communities. It turned out that the CPM method was not a proper choice for sparse networks, as it left out nodes that did not belong to any clique. However, if a network is not so sparse, then generating overlapping communities may be a better choice, especially if the context of the data suggests overlapping communities. For example, when the nodes tend to belong to more than one community at a given time. The Infomap method, however, performs better if computational complexity is an essential factor, and computational time is limited.

The results yield that evolution chains with more community states (longer chains) provide better classification results. However, there seems to be a threshold of the number of states, which make the evolution chains too short, resulting in a lack of possibility of improving the accuracy level. 

Even over 70\% of the most prominent features were obtained from the last three community states. It means that the most recent history of the community has the highest impact on its next change. This is an extremely useful conclusion if one has limited computational capabilities and cannot calculate community profiles for all states. Additionally, many new predictive features are proposed in this paper. In particular, some aggregations of node measures were used to compute the local and global microscopic features. Network structural measures were adopted as macroscopic features, and ratios of community measures to network measures were utilized as mesoscopic features. In general, the variations of the eigenvector-, eccentricity-, and closeness-based features were present in most of the selective rankings, which suggests that centrality- and distance-based measures obtained on the node level are the most valuable features.

The GEP method flexibility enabled us to investigate some other interesting scenarios, i.e., (1) adapting the transfer learning technique to the group evolution prediction problem, (2) enriching the classification model with evolution chains from a different data set, (3) appropriate selection of the observation time span to reduce the concept drift effect. All of them appeared to be quite successful.

Even though the GEP method is a flexible, generic framework, it is competitive with other approaches often dedicated to a specific problem or data set.

\section*{Supporting information}

% Include only the SI item label in the paragraph heading. Use the \nameref{SI_File} command to cite SI items in the text.
\paragraph*{S1 File.}
\label{SI_File}
{\bf Supporting information file.} Contains additional results and discussion.

\section*{Authors Contributions}
{\bf Conceptualization:} SS, PB, PK.
\newline {\bf Data Curation:} SS.
\newline {\bf Formal Analysis:} SS, MK.
\newline {\bf Funding Acquisition:} PK.
\newline {\bf Investigation:} SS, MK.
\newline {\bf Methodology:} SS, PB, MK, PK.
\newline {\bf Project Administration:} SS.
\newline {\bf Software:} SS, MK.
\newline {\bf Supervision:} PB, PK.
\newline {\bf Writing – Original Draft Preparation:} SS, PB, MK, PK.
\newline {\bf Writing – Review \& Editing:} SS, PB, PK.

\section*{Acknowledgements}
This work was partially supported by The Polish National Science Centre, the projects no. 2016/21/D/ST6/02408 and 2016/21/B/ST6/01463; by the European Union’s Horizon 2020 research and innovation programme under the Marie Skłodowska-Curie grant agreement No. 691152 (RENOIR); and by the Polish Ministry of Science and Higher Education fund for supporting internationally co-financed projects in 2016-2019, agreement no. 3628/H2020/2016/2. This research was supported in part by PLGrid Infrastructure.

\nolinenumbers

\end{document}

\vspace*{0.2in}

% Title must be 250 characters or less.
\begin{flushleft}
{\Large
\textbf\newline{Analysis of group evolution prediction in complex networks - supplementary information} % Please use "sentence case" for title and headings (capitalize only the first word in a title (or heading), the first word in a subtitle (or subheading), and any proper nouns).
}
\newline
% Insert author names, affiliations and corresponding author email (do not include titles, positions, or degrees).
\\
Stanisław Saganowski\textsuperscript{1,*},
Piotr Bródka\textsuperscript{1},
Michał Koziarski\textsuperscript{2},
Przemysław Kazienko\textsuperscript{1}
\\
\bigskip
\textbf{1} Department of Computational Intelligence, Faculty of Computer Science and Management, Wrocław University of Science and Technology, Wrocław, Poland
\\
\textbf{2} Department of Electronics, Faculty of Computer Science, Electronics and Telecommunications, AGH University of Science and Technology, Kraków, Poland
\\
\bigskip

% Use the asterisk to denote corresponding authorship and provide email address in note below.
* stanislaw.saganowski@pwr.edu.pl

\end{flushleft}

%\linenumbers
\section*{Example of a social group on Facebook}

The Facebook platform allows to perform various social activities like discussion in groups, content sharing, commenting, expressing opinions and emotions. One of the platform's tools allows to create and join independent discussion groups devoted to a specific topic. For example, there are groups intended for mothers living in Singapore, which purpose is to talk about and comment on new products for babies, share general advices about raising children, sell used clothes, etc.

By obtaining and processing data of a single discussion group we are able to create its social network graph, and furthermore, we can track its evolution. Depending on what we are trying to achieve, we would process data in a different way. In the simplest case we can assume that one post (content posted to the discussion group) and all interactions to this post (likes, comments, shares) reflect a social group at a particular time. By obtaining social groups for each post we can create a temporal social network of the considered discussion group and analyze its activeness over time. In a more complex scenario, we can analyze the content of the comments and types of interactions within each post to discover two or more groups with different opinions, e.g. recommending and criticizing a new product for babies.

% metody GEP

\section*{Group Evolution Prediction methods}
The summary of the most relevant methods for group evolution prediction known from the literature confronted with GEP, which is described in this paper, can be found in Tab.~\ref{tab:methods}.

\begin{table}[!ht]
\begin{adjustwidth}{-2.25in}{0in}
\centering
\caption{Methods for group evolution prediction.}
\begin{tabular}{p{2,2cm}|p{1,3cm}p{1,3cm}p{1,6cm}p{1,5cm}p{1,5cm}p{1,5cm}p{4cm}}
\hline

Method name & Time window type & Type of communities & Community evolution tracking method & No. of predictive features per group state & No. of classifiers analyzed & No. of real-world data sets analyzed & Prediction goal \\ \hline
GEP & any & any & any & 91 & 15 & 15 & next event (6 classes), several forthcoming events, community measure \\ \hline
İlhan et al.\cite{Ilhan:2013,Ilhan:2016} & increasing & disjoint & included in the method & 9 & 10 & 4 + 40 synthetic & next event (6 classes) \\ \hline
Takaffoli et al.\cite{Takaffoli:2014} & disjoint & disjoint & MODEC & 33 & 9 & 2 & next event (3 classes), size, cohesion \\ \hline
Diakidis et al.\cite{Diakidis:2015} & overlapping & overlapping & GED & 10 & 7 & 1 & next event (4 classes) \\ \hline
Goldberg et al.\cite{Goldberg:2011,Goldberg:2012} & disjoint & overlapping & included in the method & 20 (average) & 1 & 2 & length of community lifetime \\ \hline
Kairam et al.\cite{kairam2012life} & disjoint & unknown & unknown & 8 & 1 & 1 & community growth rate and longevity (binary classification) \\ \hline
\end{tabular}
\label{tab:methods}
\end{adjustwidth}
\end{table}

% użyte zbiory danych

\section*{Data sets used}

Fifteen real-world data sets were analyzed in the iterative process of evaluating and improving the GEP method. Nonetheless, the results presented in this work refer to ten out of fifteen analyzed data sets. The limitation was made to keep the paper clear and concise. The data sets were selected in such way, that the networks created from them had diverse characteristics, see Tab.~\ref{tab:datasets}.
During the experimental studies, the parameters of the GEP method and its components (algorithms, methods, tools) were adjusted based on the literature review, authors suggestion, previous results and experience, and sometimes as a result of repeating the experimental study endless number of times.

\begin{table}[!ht]
\begin{adjustwidth}{-2.25in}{0in}
\centering
\caption{Characteristics of the data sets used in the research}
\begin{tabular}{p{2cm}|p{0,8cm}p{1,5cm}p{1,5cm}p{1cm}p{1,2cm}p{1cm}p{5cm}}
\hline

Data set name & Source & Nodes & Edges & Avg. degree & Time span & Directed & Short description\\ \hline
DBLP & \cite{ley2002dblp} & 1,314,050 & 18,986,618 & 28.9 & 20 years & no & co-authorship of articles from the DBLP bibliography\\ \hline
Digg & \cite{de2009social} & 30,398 & 87,627 & 5.8 & 15 days & yes & replies between users on the Digg online platform\\ \hline
Facebook & \cite{viswanath2009evolution} & 46,952 & 876,993 & 37.4 & 1 year & yes & posts to other user’s wall on the Facebook social platform\\ \hline
Infectious & \cite{isella2011s} & 410 & 17,298 & 84.4 & 8 hours & no & face-to-face contacts during an exhibition\\ \hline
IrvineForum & \cite{opsahl2009clustering, rossi2015network} & 899 & 33,700 & 74.0 & 164 days & no & students activity on the UC Irvine discussion forum\\ \hline
IrvineMessages & \cite{opsahl2009clustering} & 1,899 & 59,835 & 63.0 & 6 months & yes & private messages between students of the UC Irvine\\ \hline
Loans & \cite{loans2017} & 89,269 & 3,394,979 & 76.1 & 1 year & yes & loans between users of the prosper.com platform\\ \hline
MIT & \cite{eagle2006reality} & 96 & 1,086,404 & 22.6 & 9 months & no & face-to-face contacts between students of the MIT\\ \hline
Slashdot & \cite{gomez2008statistical} & 51,083 & 140,778 & 5.5 & 32 months & yes & replies between users on the Slashdot online portal\\ \hline
Twitter & \cite{rossi2015network, conover2011political} & 18,500 & 61,200 & 6.0 & 48 days & no & retweets between users on the Twitter social platform\\ \hline

\end{tabular}
\label{tab:datasets}
\end{adjustwidth}
\end{table}

% duplicated chains

\section*{Evolution chain duplication}
\label{sec:supp:chain_duplicates}

\begin{figure}[ht]
\centering
\includegraphics[width=\linewidth]{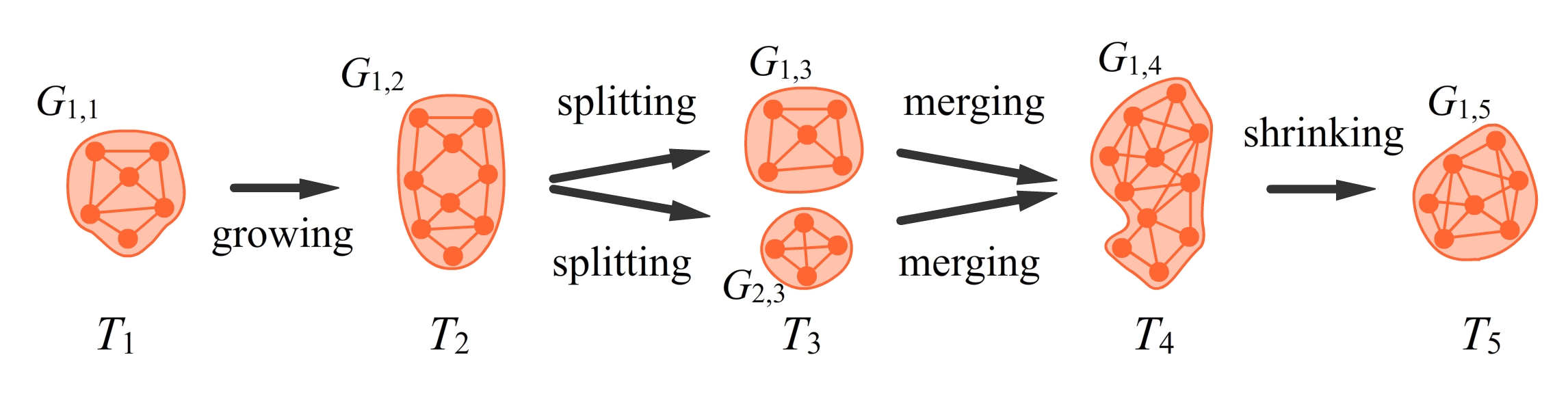}
\caption{An example of community evolution containing five states and four events.}
\label{fig:duplicated_chains}
\end{figure}

Let's consider creating evolution chains of length 2 for the exemplary community evolution depicted in Fig.~\ref{fig:duplicated_chains}. The list of evolution chains would contain five unique pairs of following states, see Tab.~\ref{tab:duplicated_2states}. As one can observe, some evolution chains are partially duplicated, e.g., state $ST_2$ and event $EV_2$ of chain $EC_1$ are the same as state $ST_1$ and event $EV_1$ of chain $EC_2$, chains $EC_2$ and $EC_3$ have the same state $ST_1$ and event $EV_1$, chains $EC_4$ and $EC_5$ share the same state $ST_2$ and event $EV_2$, and so on. The number of duplicated states and events would be even higher for a longer evolution chains. The partial duplication is a result of mixing (crossing) lifetimes of several groups, as the splitting and merging events involve at least two communities from the same time window: $G_{1,3}$ and $G_{2,3}$ in this example.

Even partially duplicated chains might be a problem, as they may affect the classification results. For example, if chains $EC_2$ and $EC_4$ would be in the training set, used to learn a classifier, and chains $EC_3$ and $EC_5$ would be in the test set, used to evaluate the classification model, the classification accuracy for chains $EC_3$ and $EC_5$ could be falsely improved, because the classifier might assign the correct event type based on “remembering” the data, rather than learning from them. One may try to remove the partially duplicated chains by applying procedure similar to the “group by” SQL command. However, this will always result in losing some information as well. In this example, grouping chains on state $ST_2$ and event $EV_2$ would result in removing chain $EC_5$.

A better solution is to use single-state chains, see Tab.~\ref{tab:duplicated_1state} for the set of chains obtained from the considered exemplary evolution. Single-state chains can also contain duplicated states, e.g., when a method for tracking evolution will assign two different event types to the same community, but it is a rare case, and such duplication can be easily removed. Throughout this paper, the process of removing partially duplicated chains is called “removing duplicates.”

\begin{table}[!ht]
\begin{adjustwidth}{-2.25in}{0in}
\centering
\caption{Evolution chains of length 2 created from the community evolution presented in Fig.~\ref{fig:duplicated_chains}.}
\begin{tabular}{c|cccc}
\hline

Evolution chain & Group state $ST_1$ in $T_i$ & Event type $EV_1$ & Group state $ST_2$ in $T_{i+1}$ & Event type $EV_2$
 \\ \hline
 
$EC_1$ & $G_{1,1}$ & growing & $G_{1,2}$ & splitting \\
$EC_2$ & $G_{1,2}$ & splitting & $G_{1,3}$ & merging \\
$EC_3$ & $G_{1,2}$ & splitting & $G_{2,3}$ & merging \\
$EC_4$ & $G_{1,3}$ & merging & $G_{1,4}$ & shrinking \\
$EC_5$ & $G_{2,3}$ & merging & $G_{1,4}$ & shrinking \\ \hline

\end{tabular}
\label{tab:duplicated_2states}
\end{adjustwidth}
\end{table}

\begin{table}[!ht]
\begin{adjustwidth}{-2.25in}{0in}
\centering
\caption{Single-state evolution chains created from the community evolution presented in Fig.~\ref{fig:duplicated_chains}.}
\begin{tabular}{c|cc}
\hline

Evolution chain & Group state $ST_1$ in $T_i$ & Event type $EV_1$  \\ \hline
$EC_1$ & $G_{1,1}$ & growing \\
$EC_2$ & $G_{1,2}$ & splitting \\
$EC_3$ & $G_{1,3}$ & merging \\
$EC_4$ & $G_{2,3}$ & merging \\
$EC_5$ & $G_{1,4}$ & shrinking \\ \hline

\end{tabular}
\label{tab:duplicated_1state}
\end{adjustwidth}
\end{table}

% ranking cech

\section*{Feature selection}
\label{sec:supp:feature_selection}

Feature extraction is an essential step that needs to be performed prior to the classification. Various measures can be used to represent the characteristic of the community at any given time step. Much effort has been made by various researchers to propose such measures, leading to their abundance. However, the high number of features is not always beneficial in the classification process. It can lead to the necessity of obtaining more data for training, which is not always feasible. Not all classifiers are resilient to the presence of uninformative features, which can weaken their performance. Finally, feature extraction can be a time consuming procedure, during both training and evaluation of the model. Due to abovementioned factors, the number of utilized features should ideally be kept to the minimum, as long as it does not lead to loss in performance.

To address this issue, feature selection process \cite{guyon2003introduction} can be performed prior to classification. Feature selection is a procedure of automatically selecting a subset of features from the larger set, possibly containing irrelevant or mutually redundant features. The aim of such task is twofold: to improve the performance of the trained model, as well as to reduce the evaluation time during its testing. However, feature selection does not address the issue of long training time. On the contrary, based on the chosen method of selection, training can be significantly prolonged. Furthermore, feature selection might itself require large amounts of data to lead to meaningful results, instead of overfitting to the task at hand. Finally, feature selection by itself gives little insight into the problem. Selected features may or may not generalize well to the other, related problems, which in uncertain when the selection is performed on a single dataset.

In the experiment described in this section, we implemented a slightly different task -- feature ranking, with the aim of providing more insight about all considered measures. Given a large number of benchmark datasets, we tried to evaluate which measures lead to the best performance during the classification. To this end, we performed a feature selection based on the evolutionary algorithm \cite{back2000evolutionary}. This procedure was repeated for various datasets and random data partitionings. Finally, we constructed a feature rankings based on the frequency of the occurrence of the given feature in the final selection. Because the feature selection strategy aims to optimize the classification performance, we postulate that the produced rankings indicate the quality of the features in the group evolution prediction task, with the quality being defined as an expected performance on the problems from the same domain. To the best of our knowledge, such evaluation has not been done before in the context of social group prediction. In the remainder of this section, we describe the proposed method more in-depth along with the most significant results.

\subsection*{Method}

The goal of the feature selection procedure is selecting a subset of features maximizing classification performance, at the same time minimizing the cost (most often computational) of producing the final subset. Given specific performance and cost measures, as well as the weights associated with both of these factors, in principle, it is possible to find the optimal feature subset, at least with regard to the available data. However, individually valuable features, i.e. the ones leading to the highest performance if used as the only predictor, will not necessarily be a part of the optimal subset. It has been shown\cite{guyon2003introduction} that the feature useless by itself can improve performance significantly when taken with others, and that the presence of highly correlated features can negatively affect the performance. Therefore, if finding the optimal feature subspace was possible, a distinction between high-quality features (those included in the optimal subset) and low-quality features (the remaining ones) can be made. However, as the number of the available features grows larger, evaluating all of the possible subsets becomes infeasible. Instead of the optimal feature subset, one has to rely on its approximation produced by the feature selection procedure. If numerous such approximations can be produced, one can associate feature quality with the frequency of the occurrence of the feature in the selected subset. Similarly, the optimality can be discussed only with regard to the available data, which is only an approximation of the underlying distribution. By selecting different data sample, we obtain a different feature subset, which is only an approximation of the optimal one.

We propose associating individual feature quality with the fraction of time it appears in the selected feature subset. On the data level, we provide diversity in the produced feature subsets by performing $5 \times 2$-fold partitioning \cite{alpaydm1999combined} on the original dataset. Furthermore, we perform a non-deterministic feature selection using basic evolutionary algorithm \cite{yang1998feature} and repeat it multiple times with a random initialization. The goal of the evolutionary algorithm is selecting a feature subset optimizing the defined fitness function.

Let us denote the original data by a tuple $(X, y)$, with $X$ being a $n \times d$ dimensional matrix of $n$ observations consisting of $d$ features each, and $y$ being a vector of $n$ class labels associated with observations. Furthermore, let us denote a $d$-dimensional binary mask encoding which features are present in the selected subset by $\hat{s}$, with $\hat{s}_i$ indicating the presence of the $i$th feature. Finally, let us denote by $X^{(\hat{s})}$ the subselection of the observations, consisting only of the features encoded in $\hat{s}$. Given the partitioning of $(X, y)$ into the training data $(X_{train}, y_{train})$, validation data $(X_{val}, y_{val})$ and test data $(X_{test}, y_{test})$, we denote the weighted $F_1$ score obtained by training the classifier on subselection $(X^{(\hat{s})}_{train}, y_{train})$ and evaluating its performance on subselection $(X^{(\hat{s})}_{val}, y_{val})$ as $F_1(\hat{s})$. We can then define the final fitness function, optimized by the evolutionary algorithm, as
\begin{equation}
f(\hat{s}) = \gamma \times F_1(\hat{s}) - \delta \times \frac{\sum_{i = 1}^{d}{\hat{s}_i}}{d},
\end{equation}
with $\gamma$ being the coefficient assigned to the classification performance, and $\delta$ -- the coefficient assigned to the number of the selected features. The evolutionary algorithm using such fitness function performs a multi-objective optimization, with the objectives: maximize the classification performance and minimize the number of selected features, and the weight associated to the objectives based on the choice of $\gamma$ and $\delta$.

For the experiments, the values of $\gamma$ and $\delta$ have been set to 0.8 and 0.2, respectively. They were chosen to keep the number of features in a given selection relatively small, with the exact value dependent on the dataset. The Random Forest was chosen as the classifier used to evaluate the classification performance of a given feature subset. The original data has been split into the training, validation and test partitions in the proportion of 0.375, 0.125 and 0.5, respectively. Finally, the following parameters of the evolutionary algorithm have been used: number of generations of 100, population size of 500, probability of mutation of 0.02, probability of crossover of 0.7, and the tournament selection with the size of 3. For each of the $5 \times 2$ folds, the evolutionary algorithm has been run 100 times, leading to 1000 feature subsets, based on which the final feature rankings have been computed.

During the conducted experimental study, all GEP features (Tab.~\ref{tab:predictiveFeatures}), and additionally the features proposed by İlhan et al. in \cite{Ilhan:2016} were analyzed. The features were obtained from 7 real-world data sets: Digg, Facebook, Infectious, IrvineMessages, Loans, MIT, Slashdot, see Tab.~\ref{tab:datasets}. The Infomap method was applied to obtain the disjoint communities, which evolution was then tracked by means of the GED method with the alpha and beta parameters set to 50\%. Time windows of various type and size, as well as the evolution chains of various length, were used to evaluate abovementioned data sets, which led to 28 separate rankings. See Tab.~\ref{tab:parameters} for the detailed information about  the data setup parameters. For each configuration from Tab.~\ref{tab:parameters}, a separate ranking was created. However, to draw more general conclusions some rankings were merged together by averaging occurrences of features in separate rankings. Only rankings containing the same set of features can be merged, thus, the same length of the evolution chain is required. Therefore, the merged rankings were obtained from the evolution chains of the following lengths: all 1-state evolution chains - Tab.~\ref{tab:topTenFeatures1s} (ids 1-12 in Tab.~\ref{tab:parameters}), all 2-state evolution chains - Tab.~\ref{tab:topTenFeatures2s} (ids 13-18 in Tab.~\ref{tab:parameters}), all 3-state evolution chains - Tab.~\ref{tab:topTenFeatures3s} (ids 19-24 in Tab.~\ref{tab:parameters}), and all 9-state evolution chains - Tab.~\ref{tab:topTenFeatures9s} (ids 26, 27, 28 in Tab.~\ref{tab:parameters}).

\begin{table}[!ht]
\begin{adjustwidth}{-2.25in}{0in}
\centering
\caption{The configuration of parameters utilized to obtain 28 feature quality rankings}
\begin{tabular}{c|ccccc}
\hline

Ranking id & Evolution chain length & Time window type & Data set & Time window size & No. of time windows \\ \hline
1 & 1 state & disjoint & Infectious & s=15min & 32 \\ 
2 &  & overlapping & Facebook & s=28days; o=14days & 27 \\ 
3 &  & overlapping & IrvineMessages & s=2days; o=1day & 192 \\ 
4 &  & overlapping & IrvineMessages & s=7days; o=3days & 47 \\ 
5 &  & overlapping & IrvineMessages & s=14days; o=7days & 26 \\ 
6 &  & overlapping & IrvineMessages & s=28days; o=14days & 12 \\ 
7 &  & overlapping & Loans & s=30days; o=15days & 23 \\ 
8 &  & overlapping & MIT & s=7days; o=3days & 57 \\ 
9 &  & overlapping & MIT & s=30days; o=15days & 14 \\ 
10 &  & increasing & Digg & s=2days & 10 \\ 
11 &  & increasing & MIT & s=30days & 10 \\ 
12 &  & increasing & Slashdot & s=36days & 10 \\ \hline
13 & 2 states & disjoint & Infectious & s=15min & 32 \\ 
14 &  & overlapping & IrvineMessages & s=7days; o=3days & 47 \\ 
15 &  & overlapping & IrvineMessages & s=14days; o=7days & 26 \\ 
16 &  & overlapping & IrvineMessages & s=28days; o=14days & 12 \\ 
17 &  & overlapping & Loans & s=30days; o=15days & 23 \\ 
18 &  & overlapping & MIT & s=30days; o=15days & 14 \\ \hline
19 & 3 states & overlapping & Facebook & s=28days; o=14days & 27 \\ 
20 &  & overlapping & IrvineMessages & s=2days; o=1day & 192 \\ 
21 &  & overlapping & MIT & s=7days; o=3days & 57 \\ 
22 &  & increasing & Digg & s=2days & 10 \\ 
23 &  & increasing & MIT & s=30days & 10 \\ 
24 &  & increasing & Slashdot & s=36days & 10 \\ \hline
25 & 8 states & increasing & Digg & s=2days & 10 \\ \hline
26 & 9 states & overlapping & Facebook & s=28days; o=14days & 27 \\ 
27 &  & overlapping & MIT & s=7days; o=3days & 57 \\ 
28 &  & increasing & Slashdot & s=36days & 10 \\ \hline

\end{tabular}
\label{tab:parameters}
\end{adjustwidth}
\end{table}

\begin{table}[!ht]
\begin{adjustwidth}{-2.25in}{0in}
\centering
\caption{The top ten features of the merged rankings for all 1-state evolution chains (ids 1-12 in Tab.~\ref{tab:parameters}). Bolded features are newly proposed.}
\begin{tabular}{c|lcl}
\hline
Rank & Feature & Occurrences & Feature type \\ \hline
1 & IlhanAging & 258 & microscopic \\
2 & network\_density & 242 & macroscopic \\
3 & \textbf{network\_leadership} & 216 & macroscopic \\
4 & \textbf{avg\_group\_eccentricity} & 185 & microscopic local \\
5 & \textbf{sum\_network\_closeness} & 178 & microscopic global \\
6 & \textbf{min\_group\_eigenvector} & 164 & microscopic local \\
7 & \textbf{max\_network\_degree\_total} & 162 & microscopic global \\
8 & \textbf{network\_reciprocity} & 146 & macroscopic \\
9 & \textbf{max\_group\_closeness} & 146 & microscopic local \\
10 & \textbf{max\_network\_closeness} & 145 & microscopic global \\ \hline
\end{tabular}
\label{tab:topTenFeatures1s}
\end{adjustwidth}
\end{table}

\begin{table}[!ht]
\begin{adjustwidth}{-2.25in}{0in}
\centering
\caption{The top ten features of the merged rankings for all 2-state evolution chains (ids 13-18 in Tab.~\ref{tab:parameters}). Bolded features are newly proposed.}
\begin{tabular}{c|lcl}
\hline
Rank & Feature & Occurrences & Feature type \\ \hline
1 & \textbf{avg\_group\_eccentricity $T_{n-1}$} & 312 & microscopic local \\
2 & \textbf{beta $T_{n-1}$} & 293 & mesoscopic \\
3 & \textbf{alpha $T_{n-1}$} & 261 & mesoscopic \\
4 & group\_coefficient\_global $T_{n-1}$ & 259 & mesoscopic \\
5 & \textbf{network\_ratio\_coefficient\_global $T_{n-1}$} & 258 & mesoscopic \\
6 & \textbf{sum\_group\_closeness $T_{n-1}$} & 251 & microscopic local \\
7 & \textbf{sum\_group\_betweenness $T_{n-1}$} & 243 & microscopic local \\
8 & network\_density $T_{n-1}$ & 242 & macroscopic \\
9 & \textbf{avg\_group\_closeness $T_{n-1}$} & 222 & microscopic local \\
10 & \textbf{max\_group\_closeness $T_{n-1}$} & 218 & microscopic local \\ \hline
\end{tabular}
\label{tab:topTenFeatures2s}
\end{adjustwidth}
\end{table}

\begin{table}[!ht]
\begin{adjustwidth}{-2.25in}{0in}
\centering
\caption{The top ten features of the merged rankings for all 3-state evolution chains (ids 19-24 in Tab.~\ref{tab:parameters}). Bolded features are newly proposed.}
\begin{tabular}{c|lcl}
\hline
Rank & Feature & Occurrences & Feature type \\ \hline
1 & \textbf{beta $T_{n-1}$} & 590 & mesoscopic \\
2 & \textbf{avg\_group\_eccentricity $T_{n-1}$} & 403 & microscopic local \\
3 & \textbf{min\_group\_eigenvector $T_{n-1}$} & 374 & microscopic local \\
4 & \textbf{beta $T_{n-2}$} & 344 & mesoscopic \\
5 & \textbf{network\_ratio\_eccentricity $T_{n-1}$} & 342 & mesoscopic \\
6 & \textbf{avg\_network\_degree\_total $T_{n-1}$} & 322 & microscopic global \\
7 & \textbf{avg\_group\_eigenvector $T_{n-1}$} & 307 & microscopic local \\
8 & IlhanInter $T_{n-1}$ & 300 & microscopic local \\
9 & \textbf{avg\_network\_degree\_in $T_{n-1}$} & 290 & microscopic global \\
10 & \textbf{max\_group\_closeness $T_{n-1}$} & 287 & microscopic local \\ \hline
\end{tabular}
\label{tab:topTenFeatures3s}
\end{adjustwidth}
\end{table}

\begin{table}[!ht]
\begin{adjustwidth}{-2.25in}{0in}
\centering
\caption{The top ten features of the merged rankings for all 9-state evolution chains (ids 26, 27, 28 in Tab.~\ref{tab:parameters}). Bolded features are newly proposed.}
\begin{tabular}{c|lcl}
\hline
Rank & Feature & Occurrences & Feature type \\ \hline
1 & \textbf{min\_group\_eigenvector $T_{n-1}$} & 521 & microscopic local \\
2 & \textbf{network\_ratio\_eccentricity $T_{n-1}$} & 433 & mesoscopic \\
3 & \textbf{sum\_network\_eigenvector $T_{n-1}$} & 428 & microscopic global \\
4 & \textbf{avg\_network\_eigenvector $T_{n-1}$} & 427 & microscopic global \\
5 & \textbf{neighborhood\_out $T_{n-1}$} & 411 & mesoscopic \\
6 & \textbf{avg\_group\_eccentricity $T_{n-1}$} & 403 & microscopic local \\
7 & \textbf{sum\_network\_betweenness $T_{n-1}$} & 399 & microscopic global \\
8 & \textbf{neighborhood\_all $T_{n-1}$} & 398 & mesoscopic \\
9 & \textbf{neighborhood\_in $T_{n-1}$} & 395 & mesoscopic \\
10 & \textbf{avg\_network\_degree\_in $T_{n-1}$} & 395 & microscopic global \\ \hline
\end{tabular}
\label{tab:topTenFeatures9s}
\end{adjustwidth}
\end{table}

% podsumowanie rankingów
In summary, the rankings of the most prominent features were different between various data sets and types of the time window, since the characteristics of the obtained temporal social networks were different. However, it was possible to identify a few measures, which appeared more often in the top ten features of various rankings. The variations of the eigenvector-, eccentricity-, and closeness-based measures were present in most of the presented shortlisted rankings, which suggests that centrality- and distance-based measures, obtained at the node level, are more informative predictors for the classifier. Surprisingly, measures describing the community in the most straightforward way, e.g., the community size or density, did not occur in the shortlisted rankings, usually taking place in the second half of the rankings. Furthermore, the macroscopic features, especially the network density, were important only when the history of the community was very short (1-2 states). Thus, when there were more historical data available, classifiers preferred past microscopic and mesoscopic features over the recent macroscopic features. What is more, the predictive features proposed by İlhan et al. in \cite{Ilhan:2016} were ranked rather low, except the IlhanAging feature, which was the most commonly used in case of the 1-state evolution chains (Tab.~\ref{tab:topTenFeatures1s}) and was usually also among the top 30 features in other rankings.

\subsection*{Reproducibility}

Experiment described in this section has been implemented in the Python programming language. Existing implementations of the classification algorithms from scikit-learn \cite{scikit-learn} and evolutionary algorithms from DEAP \cite{DEAP_JMLR2012} have been used. Code sufficient to repeat the experiment has been made publicly available at\footnote{\url{https://github.com/michalkoziarski/SocialNetworkFeatureRanking}}, whereas the necessary data, especially its partitioning used during the experiment, has been provided at \cite{harvard_data}.

% klasyfikatory

\section*{Classifiers used in the experiments}
In this experimental study 15 different classifiers, implemented in WEKA Data Mining Software\cite{hall2009weka}, were compared in term of the average F-measure value. They were gathered into six more general types.

\subsection*{Rule classifiers}
\begin{itemize}
\item \textbf{ZeroR} is the simplest classification method, which relies on the target and ignores all predictors. ZeroR classifier simply classifies the majority category (class). Although there is no predictability power in ZeroR, it is useful for determining a baseline performance as a benchmark for other classification methods.
\item \textbf{RIPPER} (JRip) is a propositional rule learner, also called Repeated Incremental Pruning to Produce Error Reduction (RIPPER), which was proposed by Cohen \cite{cohen1995fast}.
\item \textbf{DecisionTable} builds a simple decision table majority classifier \cite{kohavi1995power}. It evaluates the feature subsets using a best-first search and can use a cross-validation for the evaluation.
\end{itemize}

\subsection*{Function classifier}
\begin{itemize}
\item \textbf{Support Vector Machine} (SVM) performs classification by finding the hyperplane that maximizes the margin between classes. The vectors (cases) that define the hyperplane are the support vectors \cite{chang2011libsvm}.
\end{itemize}

\subsection*{Tree classifiers}
\begin{itemize}
\item \textbf{REPTree} is a fast decision tree learner, which builds a decision/regression tree using the information gain/variance and prunes it using a reduced-error pruning (with backfitting). It only sorts values for the numeric attributes once, and the missing values are dealt with by splitting the corresponding instances into pieces.
\item \textbf{RandomForest} is a well-known classifier for constructing a forest of random trees \cite{breiman2001random}.
\item \textbf{C4.5} (J48) is a classic classifier generating a pruned or unpruned C4.5 decision tree \cite{quinlan2014c4}.
\item \textbf{SimpleCart} is a classifier implementing the minimal cost-complexity pruning \cite{breiman1984classification}.
\end{itemize}

\subsection*{Bayes classifiers}
\begin{itemize}
\item \textbf{NaiveBayes} is a simple classifier using estimator classes; numeric estimator precision values are chosen based on analysis of the training data \cite{john1995estimating}.
\item \textbf{BayesNet} is a factored representation of the probability distributions that generalize the naive Bayesian classifier and explicitly represent statements about independence \cite{hall2009weka}.
\end{itemize}

\subsection*{Lazy classifiers}
\begin{itemize}
\item \textbf{KNN} (IBk) is a simple algorithm that stores all available cases and classifies new cases based on a similarity measure, e.g., distance functions \cite{aha1991instance}.
\item \textbf{K*} KStar is an instance-based classifier, that is the class of a test instance is based upon the class of those training instances similar to it, as determined by some similarity function. It differs from other instance-based learners in that it uses an entropy-based distance function \cite{cleary1995k}.
\end{itemize}

\subsection*{Meta-classifiers}
\begin{itemize}
\item \textbf{AdaBoost} (with DecisionStump) is a classifier for boosting a nominal class classifier using the Adaboost M1 method \cite{freund1996experiments}. DecisionStump \cite{iba1992induction} performs the classification based on entropy; missing values are treated as a separate value.
\item \textbf{Bagging} (with REPTree) bags a classifier to reduce the variance. Can do classification and regression depending on the base learner \cite{breiman1996bagging}.
\item \textbf{MultiClassClassifier} (with Logistic) is a meta-classifier for handling multi-class data sets with 2-class classifiers. This classifier is also capable of applying error correcting output codes for increased accuracy. Logistic is a classifier building a multinomial logistic regression model with a ridge estimator \cite{le1992ridge}.
\end{itemize}

\subsection*{Statistical tests of classifiers}
In order to statistically compare classifiers the Friedman test\cite{Friedman:1937} with the Shaffer post-hoc multiple comparisons\cite{Shaffer:1986} was utilized. The non-parametric statistical analysis was computed with the KEEL software tool\cite{KEEL:2009}.
The Friedman procedure was applied two times, once on the results obtained from the imbalanced data sets, and once on the results obtained from the data sets balanced with the equal size sampling technique. Tab.~\ref{tab:classifiersAvgRanks} presents the average ranks obtained by applying the Friedman procedure. The test conducted on the imbalanced data sets produced p-value=$5.25*10^{-5}$, while the test on the balanced data sets provided p-value=$1.1*10^{-3}$. Since both p-values are much lower than 0.05, the results can be considered statistically significant.

In both cases, the Bagging classifier achieved the best ranks, and the RandomForest classifier was ranked second, while the ZeroR, AdaBoost and SVM classifiers performed worst. However, the Friedman test compares only the average F-measure values obtained for all event types, it does not take into account the fact that some of the events were not classified by the particular classifier, which may be crucial if a successful classification of all event types is the goal. For instance, the Bagging classifier, which achieved the highest ranks, was not able to classify: (1) the growing event for the imbalanced Twitter data set (Fig.~\ref{fig:classifiers_comparison}A), (2) the merging event for the balanced Twitter data set, and (3) the splitting event for the imbalanced Facebook data set. At the same time, the RandomForest classifier was able to classify all event types within data sets analyzed in this experiment (Fig.~\ref{fig:classifiers_comparison}B).

However, the post-hoc comparison revealed that the difference between the Bagging and RandomForest classifiers was not statistically significant. In fact, the difference between any tree classifier and the Bagging classifier was not statistically significant. 

\begin{table}[!ht]
\begin{adjustwidth}{-2.25in}{0in}
\centering
\caption{The average ranks of classifiers obtained by applying the Friedman procedure}
\begin{tabular}{lr|lr}
\hline

\multicolumn{2}{c|}{Imbalanced data sets} & \multicolumn{2}{c}{Balanced data sets} \\
Algorithm & Avg. Ranking & Algorithm & Avg. Ranking \\ \hline
Bagging & 1.00 & Bagging & 3.00 \\
RandomForest & 3.75 & RandomForest & 4.00 \\
REPTree & 3.75 & BayesNet & 5.50 \\
C4.5 & 5.50 & DecisionTable & 5.75 \\
MultiClassClassifier & 5.50 & REPTree & 6.00 \\
DecisionTable & 6.75 & SimpleCart & 6.00 \\
K* & 6.75 & NaiveBayes & 6.75 \\
SimpleCart & 7.00 & KNN & 6.75 \\
BayesNet & 7.75 & MultiClassClassifier & 7.25 \\
KNN & 9.75 & C4.5 & 7.75 \\
RIPPER & 10.50 & K* & 9.00 \\
NaiveBayes & 10.50 & RIPPER & 10.25 \\
LibSVM & 13.00 & AdaBoost & 13.25 \\
AdaBoost & 13.50 & LibSVM & 14.00 \\
ZeroR & 15.00 & ZeroR & 14.75 \\

\end{tabular}
\label{tab:classifiersAvgRanks}
\end{adjustwidth}
\end{table}

\begin{figure}[ht]
\begin{adjustwidth}{-2.25in}{0in}
\centering
\includegraphics[width=\linewidth]{classifiers_comparison}
\caption{The classification results of different classifiers for the 1-state evolution chains obtained from the imbalanced Twitter data set. \textbf{(A)} Meta-classifiers. \textbf{(B)} Tree classifiers. \textbf{(C)} Bayes and lazy classifiers. \textbf{(D)} Rule and function classifiers.}
\label{fig:classifiers_comparison}
\end{adjustwidth}
\end{figure}

\subsection*{Adjusting classifiers parameters}
A final step in the process of applying the machine learning is tuning the classifiers’ parameters. This allows us to adjust the model to a given problem -- data set, however, it also requires a lot of efforts to run thousands of iterations slightly modifying one parameter at a time. Therefore, only the top-ranked classifiers, i.e., the Bagging, RandomForest and C4.5 classifiers, were chosen for this experimental study and only one parameter per classifier was evaluated.

\begin{figure}[ht]
\begin{adjustwidth}{-2.25in}{0in}
\centering
\includegraphics[width=\linewidth]{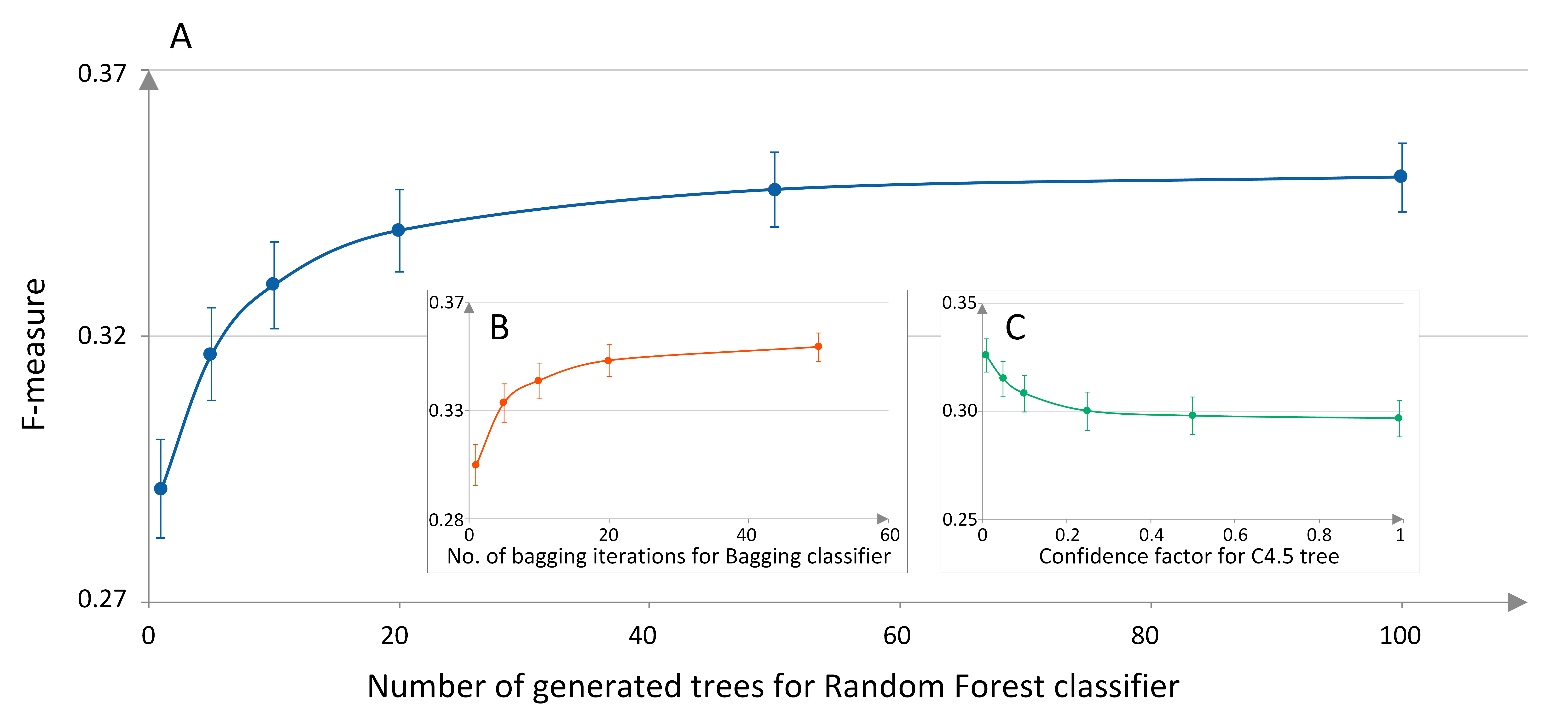}
\caption{The influence of classifiers' parameters adjustment on the F-measure value for the Facebook data set. \textbf{(A)} Tunning the number of generated trees in the Random Forest classifier. \textbf{(B)} Adjusting the number of bagging iterations in the Bagging classifier. \textbf{(C)} Fixing the confidence factor in the C4.5 tree.}
\label{fig:Adjusting classifiers parameters}
\end{adjustwidth}
\end{figure}

The result of tuning the Bagging classifier is presented in Fig. \ref{fig:Adjusting classifiers parameters}A. The correlation between the number of bagging iterations and the average F-measure demonstrates a logarithmic tendency. The average F-measure value increased substantially from 0.305 with one bagging iteration to 0.360 with 50 bagging iterations. However, the additional bagging iterations are very costly in terms of computational time. It took over 24 hours to obtain the classification results for the Bagging classifier with 50 bagging iterations and for this reason it was not further increased. Nonetheless, since the correlation between the number of bagging iterations and the average F-measure value has a logarithmic nature, it is enough to set the parameter to 10 (default value) or 20 in order to obtain a score close to the results obtained with 50 bagging iterations.

The number of generated trees by the RandomForest classifier also reveals the logarithmic correlation to the average F-measure value, Fig.~\ref{fig:Adjusting classifiers parameters}B. The overall classification score achieved with just one tree was 0.291, while the result obtained with 100 generated trees was as high as 0.350. Again, increasing the parameter value required a much longer computational time. Therefore, the experiment was discontinued for higher values. Based on the results, the parameter set to 50 seems to be a reasonable choice between the average F-measure value and the computational cost required to generate more trees. The overall score achieved with 50 trees was higher by 0.018 in comparison to the result obtained with the default parameter value (10 trees).

The confidence factor parameter of the C4.5 classifier was also correlated with the average F-measure value, see Fig.~\ref{fig:Adjusting classifiers parameters}C. Increasing the parameter value resulted in only a slight decrease of the overall score. The highest observed F-measure value was 0.326 and it was achieved with the confidence factor equal to 0.01, while the lowest overall score was 0.297, obtained with confidence of 0.99. The difference between the result achieved with the default parameter value (0.25) and the best result obtained with the parameter value set to 0.01 was 0.026.

In general, tuning the classifiers’ parameters can yield notable differences in F-measure values. Therefore, if the computational time is not limited, one may try to improve the classification results by adjusting classifiers’ parameters. In combination with other improvements, the overall gain might be very significant.

%dyskusja na temat użytej miary jakosci
\section*{Classification performance measure}
Many measures capturing the classification performance have been proposed and evaluated\cite{Cerri:2015,Ferri:2009,Sokolova:2009,Huang:2007,Yang:1999}. The most often used measures for binary classification are: accuracy, precision, recall, fscore (F-measure), specificity, and AUC (Area Under the Curve); while for multi-class classification commonly are used: average accuracy, error rate, precision, recall, and fscore (F-measure)\cite{Sokolova:2009}. The formulas for all the measures are in Eq.~\ref{eq:precision}-\ref{eq:f1-macro}.

In our study, the F-measure value (which is the harmonic mean of precision and recall) was utilized to indicate the classification performance for the particular class. Additionally, the average of all classes’ F-measure values was computed to denote the overall classification quality. The reason for using the plain average F-measure instead of the weighted F-measure, globally averaged F-measure (macro- or micro- averaged\cite{Yang:1999}), or other measures as the overall score, was to emphasize the lack of classification of some classes better. Furthermore, the plain average F-measure value considers each class to be equally important. Hence, the results of the minority classes are not lost, like in case of the accuracy measure or weighted average F-measure. What is more, the total accuracy might be sometimes misleading, e.g., when one event type suppresses others. In such cases the classifier assigns the dominating type of the event to all observations to increase the accuracy, thus, resulting in a high number of true positive and true negative classifications.

See Tab.~\ref{tab:overallMeasures} for three samples of classification results and various overall performance measures computed for these samples. The plain average F-measure has the lowest values of all measures for all three samples. Only the macro-averaged F-measure has similar values. Other measures have been impacted too much by the dominating classes and provided much higher overall scores. Sample 1 is a great example of a classifier focusing on the dominating classes. Only the plain average F-measure and the macro-averaged F-measure are reflecting the poor classification quality of the minor classes. Furthermore, the comparison between Sample 2 and Sample 3 emphasizes why the micro-averaged F-measure, weighted average F-measure, and accuracy are not considered in this paper. Both samples have an identical distribution of events and similar F-measure values, but in the case of Sample 2 the classifier was unable to classify the dissolving event. Yet, the micro-averaged F-measure, weighted average F-measure, and accuracy measures have much higher values in the case of Sample 2 than in the case of Sample 3, ignoring the missing classification. On the other hand, the plain average F-measure and the macro-averaged F-measure values indicate the unsuccessful classification of the dissolving event and have lower values in the case of Sample 2. Since the plain average F-measure value is easier to compute and understand than the macro-averaged F-measure, in this thesis, the plain average F-measure is used to represent the general classification quality.

However, any measure can be used to determine the classification performance, as long as it is appropriate to the problem the one is trying to solve.

\begin{table}[!ht]
\begin{adjustwidth}{-2.25in}{0in}
\centering
\caption{The example values of different classification performance measures showing that the average F-measure and the macro-averaged F-measure best represent the general prediction quality. Other measures have been impacted too much by the dominating classes. Sample 1 was obtained from the Digg data set, while Sample 2 and Sample 3 were obtained from the MIT data set.}
\begin{tabular}{ll|rrr|}
\cline{3-5}

 &  & Sample 1 & Sample 2 & Sample 3 \\ \hline
\multirow{7}{*}{Distribution} & Continuing & 6391 & 62 & 62 \\ 
 & Dissolving & 64 & 9 & 9 \\ 
 & Growing & 5512 & 78 & 78 \\ 
 & Merging & 504 & 29 & 29 \\ 
 & Shrinking & 4272 & 90 & 90 \\ 
 & Splitting & 235 & 38 & 38\\ \cline{2-5}
 & Sum & 16978 & 306 & 306 \\ \hline
\multirow{6}{*}{F-measure} & Continuing & 0,530 & 0,359 & 0,325 \\ 
 & Dissolving & 0.029 & 0.000 & 0.235 \\ 
 & Growing & 0.383 & 0.390 & 0.340 \\ 
 & Merging & 0.014 & 0.286 & 0.277 \\ 
 & Shrinking & 0.473 & 0.491 & 0.434 \\ 
 & Splitting & 0.015 & 0.725 & 0.709 \\ \hline
\multicolumn{2}{p{4cm}|}{Average F-measure} &  0.2406 & 0.3752 & 0.3867 \\
\multicolumn{2}{p{4cm}|}{Macro-avg F-measure} & 0.2420 & 0.3757 & 0.3876 \\
\multicolumn{2}{p{4cm}|}{Micro-avg F-measure} & 0.4430 & 0.4379 & 0.4020 \\
\multicolumn{2}{p{4cm}|}{Weighted avg F-measure} & 0.4435 & 0.4339 & 0.4014 \\
\multicolumn{2}{p{4cm}|}{Accuracy} & 0.4564 & 0.4379 & 0.4020 \\ \hline

\end{tabular}
\label{tab:overallMeasures}
\end{adjustwidth}
\end{table}

The formulas for individual quality measures are as follows:
\begin{equation} \label{eq:precision}
precision= \frac{tp}{tp+fp}
\end{equation}
where $tp$ is the number of true positive classifications and $fp$ is the number of false positive classifications;
\begin{equation} \label{eq:recall}
recall= \frac{tp}{tp+fn}
\end{equation}
where $fn$ is the number of false negative classifications;
\begin{equation} \label{eq:accuracy}
accuracy= \frac{tp+tn}{tp+tn+fp+fn}
\end{equation}
where $tn$ is the number of true negative classifications;
\mathchardef\mhyphen="2D
\begin{equation} \label{eq:fmeasure}
F \mhyphen measure=2\cdot  \frac{precision\cdot recall}{precision+recall}
\end{equation}
\begin{equation} \label{eq:micro-precision}
precision_{micro}= \frac{tp_{1}+\cdots +tp_{n}}{tp_{1}+fp_{1}+\cdots +tp_{n}+fp_{n}}
\end{equation}
\begin{equation} \label{eq:micro-recall}
recall_{micro}= \frac{tp_{1}+\cdots +tp_{n}}{tp_{1}+fn_{1}+\cdots +tp_{n}+fn_{n}}
\end{equation}
\begin{equation} \label{eq:f1-micro}
F \mhyphen measure_{micro}=2\cdot  \frac{precision_{micro}\cdot recall_{micro}}{precision_{micro}+recall_{micro}}
\end{equation}
\begin{equation} \label{eq:macro-precision}
precision_{macro}= \frac{precision_{1}+\cdots +precision_{n}}{n}
\end{equation}
\begin{equation} \label{eq:macro-recall}
recall_{macro}= \frac{recall_{1}+\cdots +recall_{n}}{n}
\end{equation}
\begin{equation} \label{eq:f1-macro}
F \mhyphen measure_{macro}=2\cdot  \frac{precision_{macro}\cdot recall_{macro}}{precision_{macro}+recall_{macro}}
\end{equation}

% tabela wszystkich cech

\section*{Predictive features}

The list of all features considered in the paper is provided in Tab.~\ref{tab:predictiveFeatures}. The new features proposed and analyzed in this paper are highlighted in bold.

\setlength\LTleft{-2in}
\setlength\LTright{0in}
\begin{longtable}{p{0.8cm}|p{4.7cm} p{11cm}}
\hline
Group & Name & Description \\ \hline
%Nodes - microscopic local - 28
\multirow{28}{*}{\rotatebox[origin=c]{90}{~Nodes - microscopic local}} & \textbf{sum\textunderscore group\textunderscore degree\textunderscore in} & The sum of indegree \cite{freeman1978centrality} of nodes belonging to the community calculated within the community. Indegree is a node measure defining the number of connections directed to the node. \\
 & \textbf{avg\textunderscore group\textunderscore degree\textunderscore in} & The average value of indegree of nodes belonging to the community calculated within the community. \\
 & \textbf{min\textunderscore group\textunderscore degree\textunderscore in} & The minimum value of indegree of nodes belonging to the community calculated within the community. \\
 & \textbf{max\textunderscore group\textunderscore degree\textunderscore in} & The maximum value of indegree of nodes belonging to the community calculated within the community. \\
 & \textbf{sum\textunderscore group\textunderscore degree\textunderscore out} & The sum of outdegree \cite{freeman1978centrality} of nodes belonging to the community calculated within the community. Outdegree is a node measure determining the number of connections outgoing from the node. \\
 & \textbf{avg\textunderscore group\textunderscore degree\textunderscore out} & The average value of outdegree of nodes belonging to the community calculated within the community. \\
 & \textbf{min\textunderscore group\textunderscore degree\textunderscore out} & The minimum value of outdegree of nodes belonging to the community calculated within the community. \\
 & \textbf{max\textunderscore group\textunderscore degree\textunderscore out} & The maximum value of outdegree of nodes belonging to the community calculated within the community. \\
 & \textbf{sum\textunderscore group\textunderscore degree\textunderscore total} & The sum of total degree of nodes belonging to the community calculated within the community. Total degree is the sum of indegree and outdegree. \\
 & \textbf{avg\textunderscore group\textunderscore degree\textunderscore total} & The average value of total degree of nodes belonging to the community calculated within the community. \\
 & \textbf{min\textunderscore group\textunderscore degree\textunderscore total} & The minimum value of total degree of nodes belonging to the community calculated within the community. \\
 & \textbf{max\textunderscore group\textunderscore degree\textunderscore total} & The maximum value of total degree of nodes belonging to the community calculated within the community. \\
 & \textbf{sum\textunderscore group\textunderscore betweenness} & The sum of betweenness \cite{freeman1978centrality} of nodes belonging to the community calculated within the community. Betweenness is a node measure describing the number of the shortest paths from all nodes to all others that pass through that node. \\
 & \textbf{avg\textunderscore group\textunderscore betweenness} & The average value of betweenness of nodes belonging to the community calculated within the community. \\
 & \textbf{min\textunderscore group\textunderscore betweenness} & The minimum value of betweenness of nodes belonging to the community calculated within the community. \\
 & \textbf{max\textunderscore group\textunderscore betweenness} & The maximum value of betweenness of nodes belonging to the community calculated within the community. \\
 & \textbf{sum\textunderscore group\textunderscore closeness} & The sum of closeness \cite{freeman1978centrality} of nodes belonging to the community calculated within the community. Closeness is a node measure defined as the inverse of the farness, which in turn, is the sum of distances to all other nodes. \\
 & \textbf{avg\textunderscore group\textunderscore closeness} & The average value of closeness of nodes belonging to the community calculated within the community. \\
 & \textbf{min\textunderscore group\textunderscore closeness} & The minimum value of closeness of nodes belonging to the community calculated within the community. \\
 & \textbf{max\textunderscore group\textunderscore closeness} & The maximum value of closeness of nodes belonging to the community calculated within the community. \\
 & \textbf{sum\textunderscore group\textunderscore eigenvector} & The sum of eigenvector \cite{bonacich1972factoring} of nodes belonging to the community calculated within the community. Eigenvector is a node measure indicating the influence of a node in the network. \\
 & \textbf{avg\textunderscore group\textunderscore eigenvector} & The average value of eigenvector of nodes belonging to the community calculated within the community. \\
 & \textbf{min\textunderscore group\textunderscore eigenvector} & The minimum value of eigenvector of nodes belonging to the community calculated within the community. \\
 & \textbf{max\textunderscore group\textunderscore eigenvector} & The maximum value of eigenvector of nodes belonging to the community calculated within the community. \\
 & \textbf{avg\textunderscore group\textunderscore eccentricity} & The average value of eccentricity \cite{harary6graph} of nodes belonging to the community calculated within the community. Eccentricity of a node is its shortest path distance from the farthest other node in the graph. \\
 & \textbf{min\textunderscore group\textunderscore eccentricity} & The minimum value of eccentricity of nodes belonging to the community calculated within the community. Also called the groups diameter. \\
 & \textbf{max\textunderscore group\textunderscore eccentricity} & The maximum value of eccentricity of nodes belonging to the community calculated within the community. \\ 
 & avg\textunderscore group\textunderscore clustering\textunderscore coefficient & The average local clustering coefficients of all the nodes in the community\cite{wasserman1994social}. \\ \hline
%% Nodes - microscopic global - 25
\multirow{25}{*}{\rotatebox[origin=c]{90}{~Nodes - microscopic global}}  & \textbf{sum\textunderscore network\textunderscore degree\textunderscore in} & The sum of indegree of nodes belonging to the community calculated within the network. \\
 & \textbf{avg\textunderscore network\textunderscore degree\textunderscore in} & The average value of indegree of nodes belonging to the community calculated within the network. \\
 & \textbf{min\textunderscore network\textunderscore degree\textunderscore in} & The minimum value of indegree of nodes belonging to the community calculated within the network. \\
 & \textbf{max\textunderscore network\textunderscore degree\textunderscore in} & The maximum value of indegree of nodes belonging to the community calculated within the network. \\
 & \textbf{sum\textunderscore network\textunderscore degree\textunderscore out} & The sum of outdegree of nodes belonging to the community calculated within the network. \\
 & \textbf{avg\textunderscore network\textunderscore degree\textunderscore out} & The average value of outdegree of nodes belonging to the community calculated within the network. \\
 & \textbf{min\textunderscore network\textunderscore degree\textunderscore out} & The minimum value of outdegree of nodes belonging to the community calculated within the network. \\
 & \textbf{max\textunderscore network\textunderscore degree\textunderscore out} & The maximum value of outdegree of nodes belonging to the community calculated within the network. \\
 & \textbf{sum\textunderscore network\textunderscore degree\textunderscore total} & The sum of total degree of nodes belonging to the community calculated within the network. \\
 & \textbf{avg\textunderscore network\textunderscore degree\textunderscore total} & The average value of total degree of nodes belonging to the community calculated within the network. \\
 & \textbf{min\textunderscore network\textunderscore degree\textunderscore total} & The minimum value of total degree of nodes belonging to the community calculated within the network. \\
 & \textbf{max\textunderscore network\textunderscore degree\textunderscore total} & The maximum value of total degree of nodes belonging to the community calculated within the network. \\
 & \textbf{sum\textunderscore network\textunderscore betweenness} & The sum of betweenness of nodes belonging to the community calculated within the network. \\
 & \textbf{avg\textunderscore network\textunderscore betweenness} & The average value of betweenness of nodes belonging to the community calculated within the network. \\
 & \textbf{min\textunderscore network\textunderscore betweenness} & The minimum value of betweenness of nodes belonging to the community calculated within the network. \\
 & \textbf{max\textunderscore network\textunderscore betweenness} & The maximum value of betweenness of nodes belonging to the community calculated within the network. \\
 & \textbf{sum\textunderscore network\textunderscore closeness} & The sum of closeness of nodes belonging to the community calculated within the network. \\
 & \textbf{avg\textunderscore network\textunderscore closeness} & The average value of closeness of nodes belonging to the community calculated within the network. \\
 & \textbf{min\textunderscore network\textunderscore closeness} & The minimum value of closeness of nodes belonging to the community calculated within the network. \\
 & \textbf{max\textunderscore network\textunderscore closeness} & The maximum value of closeness of nodes belonging to the community calculated within the network. \\
 & \textbf{sum\textunderscore network\textunderscore eigenvector} & The sum of eigenvector of nodes belonging to the community calculated within the network. \\
 & \textbf{avg\textunderscore network\textunderscore eigenvector} & The average value of eigenvector of nodes belonging to the community calculated within the network. \\
 & \textbf{min\textunderscore network\textunderscore eigenvector} & The minimum value of eigenvector of nodes belonging to the community calculated within the network. \\
 & \textbf{max\textunderscore network\textunderscore eigenvector} & The maximum value of eigenvector of nodes belonging to the community calculated within the network. \\
 & avg\textunderscore network\textunderscore clustering\textunderscore coefficient & The average of the local clustering coefficients of all the nodes in the network\cite{wasserman1994social}. \\ \hline
%Group mesoscopic
\multirow{21}{*}{\rotatebox[origin=c]{90}{~Group - mesoscopic}} & group\textunderscore size & The number of nodes in the group. \\
 & group\textunderscore density & The number of connections between nodes in the group in relation to all possible connections between them\cite{wasserman1994social}. \\
 & group\textunderscore cohesion & The vertex connectivity of the community\cite{white2001cohesiveness}. \\
 & group\textunderscore coefficient\textunderscore global  & The ratio of the triangles and the connected triples in the community\cite{wasserman1994social}. \\
 & \textbf{group\textunderscore reciprocity} & A fraction of edges that are reciprocated within the community\cite{newman2010networks}. \\
 & \textbf{group\textunderscore leadership} & A measure describing centralization in the community (the largest value is for a star network)\cite{freeman1978centrality}. \\
 & \textbf{neighborhood\textunderscore out} & The number of nodes outside the community that have incoming connection from the nodes inside the community divided by the number of nodes in the community. \\
 & \textbf{neighborhood\textunderscore in} & The number of nodes outside the community that have outgoing connection to the nodes inside the community divided by the number of nodes in the community. \\
 & \textbf{neighborhood\textunderscore all} & The number of nodes outside the community that are  connected to the nodes inside the community divided by the number of nodes in the community. \\
 & \textbf{group\textunderscore adhesion} & The minimum number of edges needed to be removed to obtain a community which is not strongly connected\cite{white2001cohesiveness}. \\
 & \textbf{alpha} & The GED inclusion measure of group $G_i$ from time window $T_n$ in group $G_j$ from $T_{n+1}$\cite{brodka2013ged}. \\
 & \textbf{beta} & The GED inclusion measure of group $G_j$ from time window $T_{n+1}$ in group $G_i$ from $T_n$\cite{brodka2013ged}. \\
 & network\textunderscore ratio\textunderscore size & The ratio of group\textunderscore size to network\textunderscore size. \\
 & network\textunderscore ratio\textunderscore density & The ratio of group\textunderscore density to network\textunderscore density. \\
 & network\textunderscore ratio\textunderscore cohesion & The ratio of group\textunderscore cohesion to network\textunderscore cohesion. \\
 & \textbf{network\textunderscore ratio\textunderscore coefficient\textunderscore global} & The ratio of group\textunderscore coefficient\textunderscore global to network\textunderscore coefficient\textunderscore global. \\
 & \textbf{network\textunderscore ratio\textunderscore coefficient\textunderscore average} & The ratio of group\textunderscore clustering\textunderscore coefficient to network\textunderscore clustering\textunderscore coefficient. \\
 & \textbf{network\textunderscore ratio\textunderscore reciprocity} & The ratio of group\textunderscore reciprocity to network\textunderscore reciprocity. \\
 & \textbf{network\textunderscore ratio\textunderscore leadership} & The ratio of group\textunderscore leadership to network\textunderscore leadership. \\
 & \textbf{network\textunderscore ratio\textunderscore eccentricity} & The ratio of avg\textunderscore group\textunderscore eccentricity to network\textunderscore avg\textunderscore eccentricity. \\
 & \textbf{network\textunderscore ratio\textunderscore adhesion} & The ratio of group\textunderscore adhesion to network\textunderscore adhesion. \\ \hline
%%% Network - macroscopic
\multirow{9}{*}{\rotatebox[origin=c]{90}{Network - macroscopic}} & network\textunderscore size & The number of nodes in the network. \\
 & network\textunderscore density & The number of connections between nodes in the network in relation to all possible connections between them. \\
 & network\textunderscore cohesion & The vertex connectivity of the network. \\
 & network\textunderscore coefficient\textunderscore global & The ratio of the triangles and the connected triples in the network. \\
 & network\textunderscore coefficient\textunderscore average & The average of the local clustering coefficients of all the nodes in the network. \\
 & \textbf{network\textunderscore reciprocity} & A fraction of edges that are reciprocated within the network. \\
 & \textbf{network\textunderscore leadership} & A measure describing centralization in the network (the largest value is for a star network). \\
 & \textbf{network\textunderscore avg\textunderscore eccentricity} & The average value of eccentricity of nodes within the network. \\
 & \textbf{network\textunderscore adhesion} & The minimum number of edges needed to be removed to obtain a graph which is not strongly connected. \\
\caption{Predictive features - newly proposed features (bolded) and features known from the literature.}
\label{tab:predictiveFeatures}
\end{longtable}

% tabela liczby zdarzeń w zależności od parametrów GED

\section*{GED}
The GED method uses the sizes and inclusion measures of two groups in the consecutive time frames to identify the event type. The alpha and beta parameters can be adjusted according to the needs. For example, to keep only evolutions between very similar groups the values of alpha and beta should be kept high, e.g., at the level of 80\%). On the other hand, sometimes the network evolves very rapidly, and the only way to track some evolutions is to lower the alpha and beta parameters e.g. to 30\%. Tab.~\ref{tab:GEDparameters} demonstrates the influence of the alpha and beta values on the number of identified events of the particular type for the IrvinaMessages data set.

%\begin{table}[!ht]

\setlength\LTleft{-1.25in}
\setlength\LTright{0in}
\begin{longtable}{cc|cccccccc}
\hline
alpha & beta & forming & dissolving & shrinking & growing & continuing & splitting & merging & total \\ \hline
10 & 10 & 362 & 350 & 424 & 351 & 97 & 217 & 164 & 1965 \\
10 & 20 & 362 & 350 & 403 & 291 & 97 & 82 & 222 & 1807 \\
10 & 30 & 362 & 350 & 378 & 280 & 93 & 66 & 233 & 1762 \\
10 & 40 & 362 & 350 & 370 & 269 & 93 & 42 & 244 & 1730 \\
10 & 50 & 362 & 350 & 347 & 260 & 94 & 35 & 253 & 1701 \\
10 & 60 & 362 & 350 & 334 & 259 & 94 & 35 & 254 & 1688 \\
10 & 70 & 362 & 350 & 322 & 258 & 94 & 38 & 255 & 1679 \\
10 & 80 & 362 & 350 & 319 & 257 & 94 & 37 & 256 & 1675 \\
10 & 90 & 362 & 350 & 319 & 257 & 94 & 37 & 256 & 1675 \\
10 & 100 & 362 & 350 & 319 & 257 & 94 & 37 & 256 & 1675 \\\hline
20 & 10 & 362 & 350 & 360 & 323 & 91 & 280 & 70 & 1836 \\
20 & 20 & 362 & 350 & 349 & 280 & 87 & 120 & 89 & 1637 \\
20 & 30 & 362 & 350 & 327 & 272 & 84 & 60 & 97 & 1552 \\
20 & 40 & 362 & 350 & 300 & 267 & 84 & 37 & 102 & 1502 \\
20 & 50 & 362 & 350 & 273 & 262 & 86 & 17 & 107 & 1457 \\
20 & 60 & 362 & 350 & 249 & 261 & 86 & 7 & 108 & 1423 \\
20 & 70 & 362 & 350 & 231 & 260 & 86 & 6 & 109 & 1404 \\
20 & 80 & 362 & 350 & 229 & 260 & 86 & 2 & 109 & 1398 \\
20 & 90 & 362 & 350 & 229 & 260 & 86 & 2 & 109 & 1398 \\
20 & 100 & 362 & 350 & 229 & 260 & 86 & 2 & 109 & 1398 \\\hline
30 & 10 & 362 & 350 & 349 & 297 & 88 & 294 & 60 & 1800 \\
30 & 20 & 362 & 350 & 342 & 261 & 78 & 129 & 40 & 1562 \\
30 & 30 & 362 & 350 & 318 & 253 & 72 & 58 & 35 & 1448 \\
30 & 40 & 362 & 350 & 278 & 251 & 71 & 40 & 37 & 1389 \\
30 & 50 & 362 & 350 & 242 & 247 & 73 & 14 & 41 & 1329 \\
30 & 60 & 362 & 350 & 205 & 247 & 73 & 7 & 41 & 1285 \\
30 & 70 & 362 & 350 & 181 & 246 & 73 & 6 & 42 & 1260 \\
30 & 80 & 362 & 350 & 175 & 246 & 73 & 2 & 42 & 1250 \\
30 & 90 & 362 & 350 & 175 & 246 & 73 & 2 & 42 & 1250 \\
30 & 100 & 362 & 350 & 175 & 246 & 73 & 2 & 42 & 1250 \\\hline
40 & 10 & 362 & 350 & 339 & 289 & 84 & 308 & 48 & 1780 \\
40 & 20 & 362 & 350 & 337 & 248 & 75 & 137 & 22 & 1531 \\
40 & 30 & 362 & 350 & 315 & 228 & 69 & 61 & 16 & 1401 \\
40 & 40 & 362 & 350 & 270 & 221 & 64 & 39 & 18 & 1324 \\
40 & 50 & 362 & 350 & 229 & 218 & 64 & 13 & 21 & 1257 \\
40 & 60 & 362 & 350 & 185 & 218 & 64 & 5 & 21 & 1205 \\
40 & 70 & 362 & 350 & 151 & 218 & 64 & 4 & 21 & 1170 \\
40 & 80 & 362 & 350 & 143 & 218 & 64 & 0 & 21 & 1158 \\
40 & 90 & 362 & 350 & 140 & 218 & 64 & 0 & 21 & 1155 \\
40 & 100 & 362 & 350 & 140 & 218 & 64 & 0 & 21 & 1155 \\\hline
50 & 10 & 362 & 350 & 345 & 280 & 83 & 303 & 52 & 1775 \\
50 & 20 & 362 & 350 & 339 & 224 & 74 & 136 & 19 & 1504 \\
50 & 30 & 362 & 350 & 315 & 191 & 67 & 62 & 16 & 1363 \\
50 & 40 & 362 & 350 & 270 & 179 & 61 & 38 & 15 & 1275 \\
50 & 50 & 362 & 350 & 223 & 175 & 52 & 13 & 16 & 1191 \\
50 & 60 & 362 & 350 & 172 & 175 & 52 & 5 & 16 & 1132 \\
50 & 70 & 362 & 350 & 131 & 175 & 52 & 4 & 16 & 1090 \\
50 & 80 & 362 & 350 & 116 & 175 & 52 & 0 & 16 & 1071 \\
50 & 90 & 362 & 350 & 110 & 175 & 52 & 0 & 16 & 1065 \\
50 & 100 & 362 & 350 & 110 & 175 & 52 & 0 & 16 & 1065 \\\hline
60 & 10 & 362 & 350 & 343 & 272 & 83 & 305 & 43 & 1758 \\
60 & 20 & 362 & 350 & 337 & 208 & 74 & 138 & 13 & 1482 \\
60 & 30 & 362 & 350 & 314 & 166 & 67 & 63 & 8 & 1330 \\
60 & 40 & 362 & 350 & 269 & 142 & 60 & 39 & 9 & 1231 \\
60 & 50 & 362 & 350 & 222 & 134 & 50 & 14 & 11 & 1143 \\
60 & 60 & 362 & 350 & 169 & 136 & 47 & 6 & 8 & 1078 \\
60 & 70 & 362 & 350 & 129 & 136 & 46 & 2 & 8 & 1033 \\
60 & 80 & 362 & 350 & 108 & 136 & 46 & 0 & 8 & 1010 \\
60 & 90 & 362 & 350 & 101 & 136 & 46 & 0 & 8 & 1003 \\
60 & 100 & 362 & 350 & 101 & 136 & 46 & 0 & 8 & 1003 \\\hline
70 & 10 & 362 & 350 & 343 & 265 & 83 & 305 & 38 & 1746 \\
70 & 20 & 362 & 350 & 337 & 193 & 74 & 138 & 7 & 1461 \\
70 & 30 & 362 & 350 & 314 & 146 & 67 & 63 & 4 & 1306 \\
70 & 40 & 362 & 350 & 269 & 113 & 60 & 39 & 6 & 1199 \\
70 & 50 & 362 & 350 & 222 & 100 & 50 & 14 & 6 & 1104 \\
70 & 60 & 362 & 350 & 169 & 101 & 47 & 6 & 4 & 1039 \\
70 & 70 & 362 & 350 & 128 & 101 & 44 & 2 & 4 & 991 \\
70 & 80 & 362 & 350 & 106 & 101 & 43 & 0 & 4 & 966 \\
70 & 90 & 362 & 350 & 99 & 101 & 43 & 0 & 4 & 959 \\
70 & 100 & 362 & 350 & 99 & 101 & 43 & 0 & 4 & 959 \\\hline
80 & 10 & 362 & 350 & 341 & 262 & 83 & 307 & 39 & 1744 \\
80 & 20 & 362 & 350 & 337 & 188 & 74 & 138 & 7 & 1456 \\
80 & 30 & 362 & 350 & 314 & 138 & 67 & 63 & 5 & 1299 \\
80 & 40 & 362 & 350 & 269 & 105 & 60 & 39 & 4 & 1189 \\
80 & 50 & 362 & 350 & 222 & 90 & 50 & 14 & 4 & 1092 \\
80 & 60 & 362 & 350 & 169 & 88 & 47 & 6 & 2 & 1024 \\
80 & 70 & 362 & 350 & 128 & 87 & 42 & 2 & 2 & 973 \\
80 & 80 & 362 & 350 & 106 & 87 & 35 & 0 & 2 & 942 \\
80 & 90 & 362 & 350 & 99 & 87 & 35 & 0 & 2 & 935 \\
80 & 100 & 362 & 350 & 99 & 87 & 35 & 0 & 2 & 935 \\\hline
90 & 10 & 362 & 350 & 341 & 261 & 83 & 307 & 40 & 1744 \\
90 & 20 & 362 & 350 & 337 & 187 & 74 & 138 & 8 & 1456 \\
90 & 30 & 362 & 350 & 314 & 137 & 67 & 63 & 6 & 1299 \\
90 & 40 & 362 & 350 & 269 & 103 & 60 & 39 & 2 & 1185 \\
90 & 50 & 362 & 350 & 222 & 86 & 50 & 14 & 2 & 1086 \\
90 & 60 & 362 & 350 & 169 & 83 & 47 & 6 & 0 & 1017 \\
90 & 70 & 362 & 350 & 128 & 81 & 42 & 2 & 0 & 965 \\
90 & 80 & 362 & 350 & 106 & 79 & 35 & 0 & 0 & 932 \\
90 & 90 & 362 & 350 & 99 & 79 & 35 & 0 & 0 & 925 \\
90 & 100 & 362 & 350 & 99 & 79 & 35 & 0 & 0 & 925 \\\hline
100 & 10 & 362 & 350 & 341 & 261 & 83 & 307 & 40 & 1744 \\
100 & 20 & 362 & 350 & 337 & 187 & 74 & 138 & 8 & 1456 \\
100 & 30 & 362 & 350 & 314 & 137 & 67 & 63 & 6 & 1299 \\
100 & 40 & 362 & 350 & 269 & 103 & 60 & 39 & 2 & 1185 \\
100 & 50 & 362 & 350 & 222 & 86 & 50 & 14 & 2 & 1086 \\
100 & 60 & 362 & 350 & 169 & 83 & 47 & 6 & 0 & 1017 \\
100 & 70 & 362 & 350 & 128 & 80 & 42 & 2 & 0 & 964 \\
100 & 80 & 362 & 350 & 106 & 78 & 35 & 0 & 0 & 931 \\
100 & 90 & 362 & 350 & 99 & 78 & 35 & 0 & 0 & 924 \\
100 & 100 & 362 & 350 & 99 & 78 & 35 & 0 & 0 & 924 \\\hline

\caption{The number of events of the particular type tracked with the GED method for different values of the alpha and beta parameters for the IrvinaMessages data set.}
\label{tab:GEDparameters}
\end{longtable}

\section*{Acknowledgements}
This work was partially supported by The Polish National Science Centre, the projects no. 2016/21/B/ST6/01463 and 2016/21/D/ST6/02408 and by the European Union’s Horizon 2020 research and innovation programme under the Marie Skłodowska-Curie grant agreement No. 691152 (RENOIR) and the Polish Ministry of Science and Higher Education fund for supporting internationally co-financed projects in 2016-2019, agreement no. 3628/H2020/2016/2. This research was supported in part by PLGrid Infrastructure.

\nolinenumbers

% Either type in your references using
% \begin{thebibliography}{}
% \bibitem{}
% Text
% \end{thebibliography}
%
% or
%
% Compile your BiBTeX database using our plos2015.bst
% style file and paste the contents of your .bbl file
% here. See http://journals.plos.org/plosone/s/latex for 
% step-by-step instructions.
% 